\def\cm{{\rm\thinspace cm}}

\def\erg{{\rm\thinspace erg}}

\def\K{{\rm\thinspace K}}
\def\keV{{\rm\thinspace keV}}
\def\km{{\rm\thinspace km}}

\def\Mpc{{\rm\thinspace Mpc}}
\def\Msun{\hbox{$\rm\thinspace M_{\odot}$}}
\def\pc{{\rm\thinspace pc}}

\def\s{{\rm\thinspace s}}
\def\yr{{\rm\thinspace yr}}

\def\mnras{MNRAS\ }
\def\apj{ApJ\ }
\def\aj{AJ\ }
\def\araa{ARAA\ }
\def\nat{Nature\ }
\def\apjl{ApJL\ }
\def\aap{Astr.Ap.\ }
\def\aapr{Astr.Ap.Rev.\ }
\def\apjs{ApJS\ }
\def\physrep{Phys.Rep.\ }
\def\nar{New Ast. Rev.\ }
\def\procspie{SPIE\ }
\def\planss{Plan.Sp.Sci.\ }

\def\cmpssq{\hbox{$\cm\s^{-2}\,$}}

\def\ergps{\hbox{$\erg\s^{-1}\,$}}

\def\kmps{\hbox{$\km\s^{-1}\,$}}

\def\Msunpyr{\hbox{$\Msun\yr^{-1}\,$}}
\def\pcm{\hbox{$\cm^{-2}\,$}}
\documentclass{article}

\usepackage{graphicx}
\usepackage{setspace}
\usepackage{mathptmx}

\begin{document}

\input epsf.tex    
\input epsf.def   

\input psfig.sty

\title{Observational Evidence of AGN Feedback}

\markboth{Observational Evidence of AGN Feedback}{A.C. Fabian}

\author{A.C Fabian\\
            {\small Institute of Astronomy, Madingley Road Cambridge CB3 0HA,
UK}}

\date{}
\maketitle

\singlespacing
\begin{abstract}
Radiation, winds and jets from the active nucleus of a massive galaxy
can interact with its interstellar medium leading to ejection or
heating of the gas. This can terminate star formation in the galaxy
and stifle accretion onto the black hole. Such Active Galactic Nucleus
(AGN) feedback can account for the observed proportionality between
central black hole and host galaxy mass. Direct observational evidence
for the radiative or quasar mode of feedback, which occurs when the
AGN is very luminous, has been difficult to obtain but is accumulating
from a few exceptional objects.  Feedback from the kinetic or radio
mode, which uses the mechanical energy of radio-emitting jets often
seen when the AGN is operating at a lower level, is common in massive
elliptical galaxies. This mode is well observed directly through X-ray
observations of the central galaxies of cool core clusters in the form
of bubbles in the hot surrounding medium. The energy flow, which is
roughly continuous, heats the hot intracluster gas and reduces
radiative cooling and subsequent star formation by an order of
magnitude. Feedback appears to maintain a long-lived heating/cooling
balance. Powerful, jetted radio outbursts may represent a further mode
of energy feedback which affect the cores of groups and subclusters.
New telescopes and instruments from the radio to X-ray bands will come
into operation over the next few years and lead to a rapid expansion
in observational data on all modes of AGN feedback.
\end{abstract}

\setcounter{tocdepth}{2}
\tableofcontents

\section{Introduction}
It has been realised over the past decade that the black hole at the
centre of a galaxy bulge is no mere ornament but may play a major role
in determining the final stellar mass of the bulge. The process by
which this occurs is known as AGN (Active Galactic Nucleus) feedback
and it takes place through an interaction between the energy and
radiation generated by accretion onto the massive black hole (the AGN)
and the gas in the host galaxy. The possibility arises where the
intense flux of photons and particles produced by the AGN sweeps the
galaxy bulge clean of interstellar gas, terminates star formation, and
through lack of fuel for accretion, terminates the AGN.

The ratio of the size of the black hole to its massive host galaxy is
tiny and similar to a coin in comparison to the Earth. The feedback
process must therefore operate over a hundred to thousand millionfold
range of scale (i.e. $10^8-10^9$). The details of the feedback are
complex and the observational evidence is not always clear.

The overall picture in terms of energetics is fairly straightforward
and at least two major modes have been identified, differentiated by
the nature of the energy outflow near the black hole. The first is the
{\it radiative} mode, also known as the quasar or wind mode, which
operates, or operated, in a typical bulge when the accreting black
hole was close to the Eddington limit. It is most concerned with
pushing cold gas about.  The second mode is the {\it kinetic} mode,
also known as the radio jet, or maintenance mode.  This typically
operates when the galaxy has a hot halo (or is at the centre of a
group or cluster of galaxies) and the accreting black hole has
powerful jets. At the present epoch it tends to occur at a lower
Eddington fraction and in more massive galaxies and involves hot gas.

A further, ill-understood, mode may be associated with giant radio
sources, which range in size up to a few Mpc. The energies in these
sources are prodigious and approach the binding energy of the gas
in groups and subclusters.  

It is easy to demonstrate that the growth of the central black hole by
accretion can have a profound effect on its host galaxy. If the
velocity dispersion of the galaxy is $\sigma$ then the binding energy
of the galaxy bulge, which is of mass $M_{\rm gal},$ is $E_{\rm
  gal}\approx M_{\rm gal} \sigma^2$. The mass of the black hole is
typically observed to be $ M_{\rm BH}\approx 1.4\times 10^{-3} M_{\rm
  gal}$ (Kormendy \& Gebhardt 2001; Merritt \& Ferrarese 2001;
H\"aring \& Rix 2004). Assuming a radiative efficiency for the
accretion process of 10\%, then the energy released by the growth of
the black hole is given by $E_{\rm BH}= 0.1 M_{\rm BH} c^2$. Therefore
$E_{\rm BH}/E_{\rm gal}\approx 1.4\times 10^{-4}(c/\sigma)^2.$ Most
galaxies have $\sigma<400\kmps,$ so $E_{\rm BH}/E_{\rm gal} >
80$. The energy produced by the growth of the black hole therefore
exceeds the binding energy by a large factor. If even a small fraction
of the energy can be transferred to the gas, then an AGN can have a
profound effect on the evolution of its host galaxy.

Fortunately accretion energy does not significantly affect the stars
already existing in the host galaxy, or there would not be any
galaxies as we know them. Nevertheless, observational evidence is
reviewed here that energy and momentum from accretion onto the central
black hole can couple strongly with the gas from which new stars
forms.

AGN feedback is a relatively young topic and a wide range of argument
and opinion has been expressed.  One opinion holds that AGN feedback
locks the mass of the black hole to that of its host galaxy bulge, and
determines the ultimate stellar mass of the bulge. Another opinion has
it as just one of many processes of comparable importance in galaxy
evolution; with the black hole -- galaxy bulge mass correlation merely
being a result of repeated mergers (e.g. Jahnke \& Macci{\'o} 2011).
There is little or no evidence at the present time for AGN feedback
operating in low mass galaxies where stellar feedback is important, or
that it significantly affects galaxy disks, or pseudobulges.

The clearest observational evidence for AGN feedback is found in the
most massive galaxies known, Brightest Cluster Galaxies (BCGs) in cool
core clusters of galaxies. Without energy input through kinetic
feedback, many BCGs would be yet more massive and appear as brilliant,
giant starbursts.

AGN feedback features in many theoretical, numerical and semi-analytic
simulations of galaxy growth and evolution (e.g. Kauffmann \& Haehnelt
2000; Granato et al 2004; Di Matteo et al 2005; Springel, Di Matteo \&
Hernquist 2005; Bower et al 2006; Croton et al 2006; Ciotti et al
2010; Hopkins et al 2006, Scannapieco et al 2011).  They are not
reviewed here. The 9 orders of magnitude in physical scale means that
all such simulations include subgrid assumptions and approximations.

The review begins with a brief outline of the physics behind the
radiative mode, then discusses the effects of radiation pressure on
dusty gas, followed by AGN winds, outflows and AGN evolution. The
radiative mode is the most likely AGN feedback explanation for the
black hole mass -- stellar velocity dispersion relation ($M-\sigma$,
see Section 2), since it relies on the accretion being radiatively
efficient and close to the Eddington limit. It was probably most
effective back at $z\sim2-3$ when quasar activity peaked and galaxies
were most gas rich. Much of the feedback action involves absorption of
the quasar radiation, which obscures the AGN itself, so direct
observational evidence is patchy at the moment.

If feedback empties a massive galaxy of gas it will then refill with
at least stellar mass loss if isolated, or with intracluster plasma if
in a cluster or group. Keeping it empty, or at least keeping the gas
hot so it does not cool, appears to be the role of the kinetic
mode, which is discussed next. This mode gives the most dramatic
observational examples of AGN feedback in terms of bubbles in the
cores of clusters. 

Energy injection from powerful giant radio galaxies is treated
last. This may have a drastic impact on the gas in groups and
subclusters. Observational evidence is poor, however, since much of
the power in relativistic electrons (but not protons) is lost in
Compton scattering of the Cosmic Microwave Background, the energy
density of which was much higher in the past. 

The review finishes by considering whether the long term behaviour of
AGN, and the modes of accretion, parallels the outbursts of Galactic,
stellar mass, binary black holes which tend to be radiatively
efficient and windy at high luminosity, and radiatively inefficient
and jetted at low luminosity. Future observational prospects for AGN
feedback, which are very bright, are treated in a concluding section.

\section{The Radiative or Wind Mode}
Silk \& Rees (1998, see also Haehnelt, Natarajan \& Rees 1998) pointed
out that a quasar at the Eddington limit can prevent accretion into a
galaxy at the maximum possible rate provided that
$$M_{\rm BH} \sim{{f\sigma^5\sigma_{\rm T}}\over{4\pi G^2 m_{\rm p} c}},$$
where $\sigma_{\rm T}$ is the Thomson cross section for electron
scattering and $f$ is the fraction of the galaxy mass in gas . The
galaxy is assumed to be isothermal with radius $r$, so that its mass
is $M_{\rm gal}=2\sigma^2 r/G.$ The maximum collapse rate, $\sim
2f\sigma^3/G$, is equivalent to the gas content, $fM_{\rm gal}$,
collapsing on a freefall time, $r/\sigma$, requiring a power of $\sim
f\sigma^5/G$ to balance it which is limited by the Eddington
luminosity $L_{\rm Edd}=4\pi G M_{\rm BH} m_{\rm p} c/\sigma_{\rm T}$.
The argument is based on energy which is necessary but may not be
sufficient for ejecting matter (the rocket equation, for example, is
based on momentum).

Momentum balance gives an expression (Fabian 1999, Fabian, Wilman \&
Crawford 2002, King 2003, 2005, Murray, Quataert \& Thompson 2005)
$$M_{\rm BH}={{f\sigma^4\sigma_{\rm T}}\over{\pi G^2 m_{\rm p}}},$$
which is about $c/\sigma$ times larger and in striking agreement with
the observed black hole mass vs stellar velocity disperson ($M_{\rm
  BH}-\sigma$) relation (e.g. G\"ultekin et al 2009) for a plausible
gas mass fraction $f\sim 0.1$.

There are several ways to derive the above formula. A simple one is to
assume that the radiation pressure from the Eddington-limited quasar
$L_{\rm Edd}/c$ has swept the gas, of mass $M_{\rm gas}=fM_{\rm gal},$
to the edge of the galaxy. Balancing the outward radiation force with
the inward one due to gravity gives
$${{4\pi GM_{\rm BH}m_{\rm p}}\over \sigma_{\rm T}}= 
{L_{\rm Edd}\over c}={{GM_{\rm gal}M_{\rm gas}}\over
  r^2}={{fGM_{\rm gal}^2}\over r^2}={{fG}\over r^2}{\left({2\sigma^2
    r}\over G\right)^2}$$
i.e.
$${{4\pi G M_{\rm BH}m_{\rm p} }\over{\sigma_{\rm T}
  }}={{f4\sigma^4 }
\over G},$$
from which the result follows. The cancellation of the radius in the
formula means it applies within the galaxy.

The agreement that this simple formula gives with the observed $M_{\rm
  BH} - \sigma$ relation can be interpreted as (weak) observational
evidence for AGN feedback.

\subsection{Radiation pressure on dust}

The interaction cannot rely on radiation pressure on electrons as in
the standard Eddington-limit formula, since if the quasar is locally
at its Eddington limit then it must be far below the Eddington limit
when the mass of the galaxy is included. Quasars appear to respect the
Eddington limit, see e.g. Kollmeier (2006) and Steinhardt \& Elvis
(2010). (King 2003 does however invoke super-Eddington luminosities.)
The interaction has to be much stronger, either due to a wind
generated close to the quasar which then flows through the galaxy
pushing the gas out, or to dust embedded in the gas, as expected for
the interstellar medium of a galaxy (Laor \& Draine 1993, Scoville \&
Norman 1995, Murray, Quartaert \& Thompson 2005). Dust grains embedded
in the gas will be partially charged in the energetic environment of a
quasar, which binds them to the surrounding partially-ionized
gas. $L_{\rm Edd}$ is reduced by a factor of $\sigma_{\rm
  d}/\sigma_{\rm T}$, where $\sigma_{\rm d}$ is the equivalent dust
cross section per proton, appropriately weighted for the dust content
of the gas and the spectrum of the quasar.

We find that $\sigma_{\rm d}/\sigma_{\rm T}$ is about 1000 for a gas
with a Galactic dust-to-gas ratio exposed to a typical quasar spectrum
(Fabian, Vasudevan \& Gandhi 2008a), dropping to 500 for low Eddington
ratio objects. This means that a quasar at the standard Eddington
limit (for ionized gas) is at the effective Eddington limit (for dusty
gas), $L_{\rm Edd}',$ of a surrounding object 1000 times more
massive. Both AGN and galaxy are then at their respective Eddington
limits. Is this just a coincidence or the underlying reason why $
M_{\rm gal}/M_{\rm BH}\sim 1000$?

AGN show indications of an effective Eddington limit in the
distribution of absorption column densities, $N_{\rm H},$ as a
function of Eddington ratio, $\lambda=L_{\rm bol}/L_{\rm Edd}$ found
in several surveys (Raimundo et al 2010).  There is a lack of objects
with column densities, $N_{\rm H},$ in the range $3\times 10^{21} -
3\times 10^{22}\pcm$ and $\lambda>0.1$. This is unlikely to be an
observational selection effect since such objects would be X-ray
bright. Any object found in that zone would be of great interest as it
could test radiative feedback on dust. The gas should be outflowing.

Interstellar gas in an AGN host evolves such that any which strays
into a region where $L_{\rm Edd}'>1$ is pushed outwarde. Gas which is
introduced to a galaxy can remain, fuelling both the black hole and
star formation, provided both $L_{\rm Edd}'$ and $L_{\rm Edd}$ remain
below unity. Repitition of this process could drive $ M_{\rm BH}/
M_{\rm gal}\rightarrow \sigma_{\rm T}/\sigma_{\rm d}=10^{-3}$.

If the repeated action of radiation pressure on dust is responsible
for the $M_{\rm BH}-\sigma$ relation then it must cause the bulge mass
to be $\sigma_{\rm d}/\sigma_{\rm T}$ times the black hole mass,
$${M_{\rm gal}}\sim {{f\sigma^4\sigma_{\rm d}}\over{\pi G^2
    m_{\rm p}}}.$$ For a constant mass-to-light ratio, this
corresponds to the Faber--Jackson (1976) relation.

Since $M_{\rm gal}=2\sigma^2 r/G$, then
$${{\sigma^2}\over r}\sim {{2\pi G m_{\rm
      p}}\over {f\sigma_{\rm d}}}.$$ Feedback should shape both the
black hole and the galaxy bulge and may even lead to some aspects
(e.g. $\sigma^2\propto r$) of the Fundamental Plane (Faber et al 1987;
Djorgovski et al 1987).  (It is curious that the above value for
${{\sigma^2}\over r}\sim10^{-8}\cmpssq$ is close to the fiducial
acceleration $a_o$ in MOND theory (Sanders \& McGaugh 2002).

Galaxies occur in dark matter haloes, which define the outer
gravitational potential well. The total mass of the halo can be an
order of magnitude more than that of the stellar part of the
galaxy. Silk \& Nusser (2010) have shown that AGN feedback may not be
energetic enough to eject all the gas from the halo, as well as the
galaxy, if the gas moves at the (local) escape velocity.

In the speculative process described above, where cycles of AGN
activity push the gas out of the galaxy, then the gas may end up
trapped in the halo. It is plausible however that the squeezing of the
gas during the ejection process triggers star formation, leading to
shells of stars on ever larger (bound) radial orbits as the galaxy
grows. This inside-out growth pattern superficially matches
observations of the development of the radii of early-type galaxies
since $z\sim 2$ (Van Dokkum et al 2010) just after quasar activity had
peaked.

\subsubsection{Optical depth effects and anisotropy}

The above discussion assumes that the infrared radiation produced by
the absorption of quasar radiation by dust is not heavily trapped. If
it is then the net radiation pressure is increased proportional to the
optical depth, and the relationships become more complicated.

The bulk of quasar radiation originates from an accretion disc
and has a bipolar radiation pattern. This both allows accretion to
proceed along the disc plane, fuelled by mergers, cold flows, or just
secular evolution of the galaxy, while at the same time pushing matter
out strongly along the disc axis. Gas in the body of the galaxy at
100~pc or more will be mostly swept up along that axis, the
gap around the equator preventing significant large-scale trapping of
the radiation.  This means that galaxies growing under strong
radiation feedback as envisaged above could appear elongated along
the radiation axis.  

\subsection{AGN winds}

If the main interaction is due to winds, not to radiation pressure,
then the wind needs to have a high column density $N$, high
velocity $v$, high covering fraction $f$, all at large radius $r$. The
kinetic luminosity of a wind is 
$${{L_{\rm w}}\over L_{\rm Edd}}={f\over 2}{r\over r_{\rm
    g}}\left({v\over c}\right)^3{N\over N_{\rm T}},$$ where $r_{\rm
  g}$ is the gravitational radius $GM/c^2$ and $N_{\rm T}=\sigma_{\rm
  T}^{-1}=1.5\times 10^{24}\pcm.$ For high wind power, $L_{\rm w}\sim
L_{\rm Edd}$, then if 
$v\sim 0.1 c$ then values of $r>10^3r_{\rm g}$ and $N\sim N_{\rm T}$
are needed. If the
wind is pressure driven then it might be expected that the gas is
accelerated where $v$ is the local escape velocity so $r\sim (c/v)^2
r_{\rm g}$.  

To produce $M_{\rm BH}\propto \sigma^4$ scaling the thrust of the wind
needs to be proportional to the Eddington limit. This seems plausible
if the wind is dusty or the acceleration is due to radiation pressure
acting on resonance lines in the gas. A problem  with a high
velocity dust-driven wind is that dust is unlikely to survive close to
the black hole where the escape velocity is high.  It is not clear
that wind strength is proportional to the Eddington limit if the
wind is accelerated magnetically by, say, the Blandford-Payne (1982)
mechanism.

The commonest way in which AGN winds are observed is by line
absorption of the quasar continuum by intervening wind material.  The
X-ray warm absorbers commonly seen in Seyfert galaxies (Reynolds 1997,
Crenshaw, Kraemer \& George 2003) flowing at $\sim 1000\kmps$ are
insufficient, by a large factor (Blustin et al 2005). Faster winds are
required, such as those seen in UV observations of BAL quasars
(e.g. Ganguly et al 2007; Weymann et al 1991) and in X-ray
observations of some AGN (e.g. Pounds et al 2003; Reeves et al 2009;
Tombesi et al 2010, Fig.~1), with velocities of tens of thousands of
$\kmps$. Establishing that the kinetic power of the wind is sufficient
has proven difficult: if the evidence of the wind is from blueshifted
absorption lines then obtaining the covering fraction and radius of
the wind requires indirect arguments. Tombesi et al (2012) estimate
that the mass outflow rate exceeds 5\% of the mass accretion rate and
that the lower limit on the kinetic power of the outflows in
individual objects ranges from $10^{42.6} - 10^{44.6}\ergps$.
  
\begin{figure}
\begin{center}
\hbox{\includegraphics[width=2.in,angle=-90]{tombesi_fig_8.ps} 
\hspace{0.1cm}
\includegraphics[width=1.5in,angle=-90]{newfig1b.ps}} 
\vspace{2cm}
\includegraphics[width=4in]{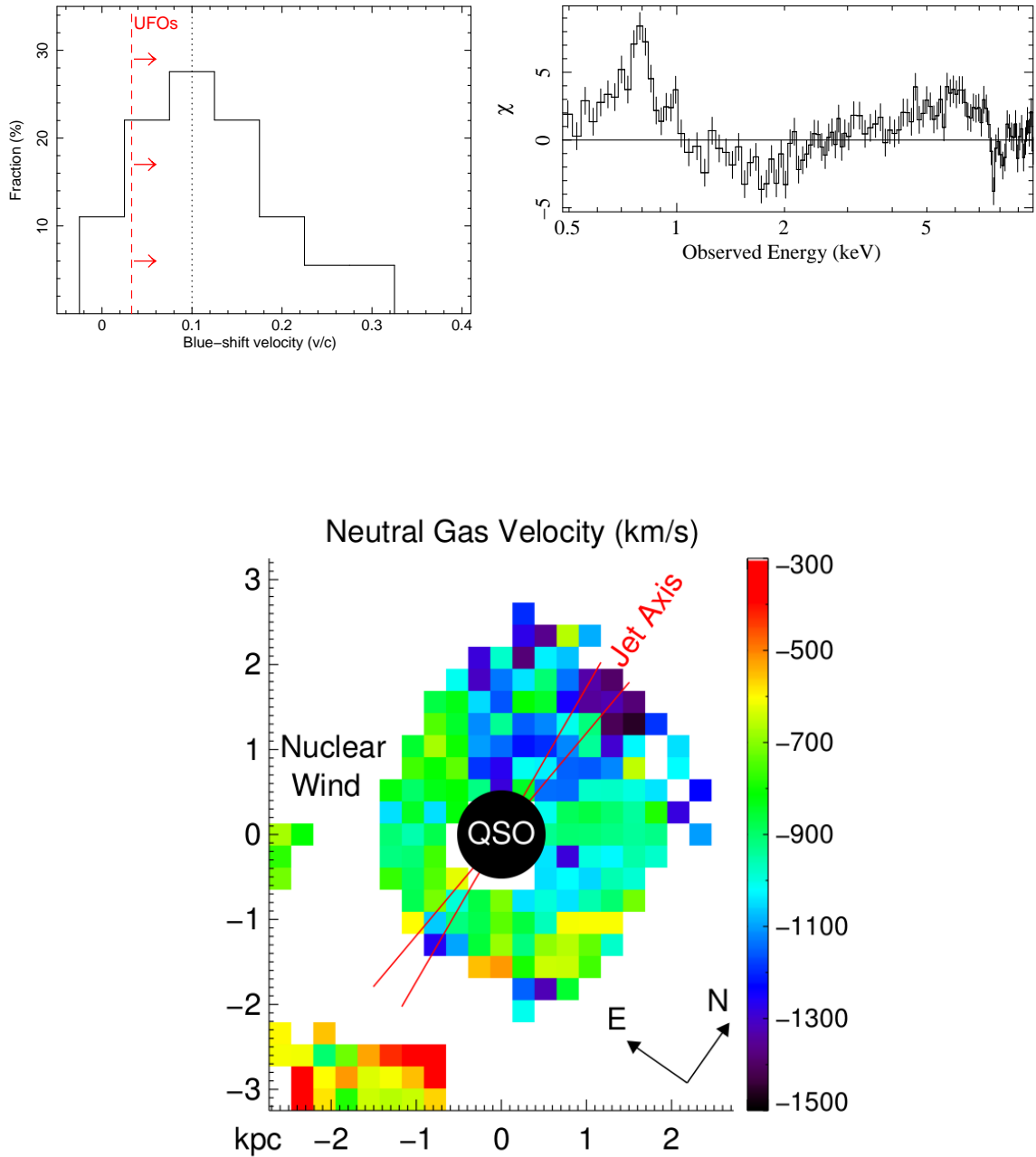}
\caption{Left: Distribution of wind velocities inferred from from
  X-ray absorption features in low redshift AGN (Tombesi et al
  2010). Right: Blueshifted X-ray absorption features in the most
  luminous low redshift quasar, PDS\,456 at $z=$  (Reeves et al
  2009). Note the 9~keV absorption feature in the rest frame of the
  quasar, presumed to be due to FeXXV (rest frame 6.7~keV)
  blueshifted by 0.3. 
Lower:
  Neutral gas velocity map in the quasar/merger object Mrk231 (Rupke
  \& Veilleux 2011). }
   \label{fig1}
\end{center}
\end{figure}

Careful work on some quasars (Dunn et al 2010, Moe
et al 2010, Saez et al 2010) has established wind powers at 5--10 per
cent of the accretion power, which is sufficient to eject gas from a
galaxy. This is backed up by a range of less direct estimates.  An
understanding of the overall effect of powerful winds also requires
estimates of their longevity.

Generally, good evidence for AGN winds occurs in unobscured AGN, where
the UV spectrum can be directly seen. There is then little cold gas
along our line of sight in the host galaxy to be swept out, either
because this has already occured or there is little cold gas in the
first place. Where significant cold gas is present in the galaxy, the
intrinsic AGN spectrum may be blocked from view and feedback is
inferred from the velocity field of any outflow in the galaxy.
  
\subsection{Galaxy outflows}

Evidence of AGN feedback is clearly seen in some galactic
outflows. Galactic winds and starburst superwinds (Veilleux, Cecil \&
Bland-Hawthorn 2005; Heckman et al 2000, Strickland \& Heckman 2009; Weiner
at al 2009) can range from tens to over a thousand Solar masses per
year with velocities of a few 100$\kmps$ for the cool components. Most
of the lower velocity winds are considered to be powered by stellar
processes such as supernovae. Identifying the effects of AGN feedback
in outflows often relies on observing higher velocity (e.g. $>
500\kmps$) components and an outflow power exceeding that predicted by
any central starburst.  The details are not easy to discern, nor is
there yet a simple clear dividing line between star- and AGN-driven
outflows. $500\kmps$ is $\sim 1\keV$ per particle and difficult to
achieve with stellar processes in large masses of cold molecular gas.  There
should of course be a powerful AGN at the centre of the galaxy. High
accretion requires a high fuelling rate which often leads to high
obscuration of the AGN. The obscuration by the surrounding gas, makes
observations of the UV and soft X-ray bands where absorption features
are most readily detected, more difficult.

An important object where both the AGN and outflow are seen is the low
redshift, $z=0.04,$ quasar/merger Mrk\,231. Rupke \& Veilleux (2011; Fig.~1)
map a strong outflow in it with a velocity of $\sim 1100\kmps$ and an
outflow rate of $420\Msunpyr$, several times greater than the star
formation rate. The outflow power is about one per cent of the
bolometric luminosity of the AGN (see also Ferruglio et al 2010;
Fischer et al 2010). 

The optical/UV spectrum of Mrk231 shows that it is a low ionization
BAL (LoBal) quasar with strong additional absorption (Smith et al
1995). A study of FeLoBAL quasars at $0.8<z<1.8$ by Farrah et al
(2012) concludes that radiatively driven outflows from AGN act to
curtail obscured star formation (inferred from the IR luminosity) in
the host galaxies of reddened quasars to less than $\sim 25$\% of the
total IR luminosity.

Sturm et al (2011) have used Herschel-PACS to observe the far-infrared
spectrum of the OH 79$\mu$m feature in several low redshift
Ultra-Luminous Infrared Galaxies (ULIRGs). They find high velocities
above $1000\kmps$ and mass outflow rates of up to $1200\Msunpyr$ in
the AGN dominant ones. The gas depletion times range from
$10^6-10^8\yr$. Their result leaves little doubt that massive outflows
are generated by AGN.

There are many recent reports of outflows from galaxies hosting AGN.
$1000\kmps$ outflows have been seen either side of an obscured quasar
at $z=0.123$ (Greene, Zakamska \& Smith 2011), in massive
post-starburst galaxies at $z\sim 0.6$ (Tremonti, Moustakas \&
Diamond-Stanic 2007), and covering 4--8~kpc of an Ultra Luminous
Infrared Galaxy hosting an AGN at $z\sim 2$ (Alexander et al 2010). A
region over a luminous quasar at $z=2.4$ shows star formation
suppressed, as inferred from decreased H$\alpha$ emission, where the
outflow velocity, deduced from [OIII] emission, is highest
(Cano-D{\'i}az et al 2011). Absorption features in the spectrum of a
background quasar shining through the halo 108~kpc out from a $z=2.4$
quasar reveal extreme kinematics in metal-rich cold gas (Prochaska \&
Hennawi 2009).

A spectacular example is the $1300 \kmps$ outflow in a redshift 6.4
quasar revealed by broad wings of the [CII] emission line (Maiolino et
al 2012). The kinetic power in the outflow is $\sim~2 \times 10^{45}
\ergps$ and the $2\times 10^{10}\Msun$ molecular gas content of the
host galaxy, inferred by CO observation, is ejected in $<10^7\yr$.

Strong outflows are also seen in radio galaxies at both low
(e.g. Morganti et al 2007) and high redshifts (Nesvadba et al
2008, 2011). The above results are a foretaste of what can be expected over the
next few years as instrumentation and techniques improve.

\subsection{From the peak to the late evolution of AGN and quasars}

An important discovery of the past decade was the cosmic downsizing of
AGN. The most luminous and massive AGN were most numerous at redshifts
of 2--2.5, the less-luminous peaked at successively lower redshifts
with the least luminous peaking around redshift one. Downsizing of AGN
was first seen in X-ray surveys (Ueda et al 2003, Hasinger, Miyaji \&
Schmidt 2005;
Barger et al 2005) where the nucleus stands out clearly above the
surrounding galaxy in even the low-luminosity objects. Later work in
optical and other bands confirms this picture. Downsizing is also see
in radio AGN (Rigby et al 2011). 

The behaviour is the opposite of what is simply predicted in a
hierarchical CDM universe, where the most massive objects (clusters of
galaxies) form last. It indicates that something is quenching quasar
behaviour and the most widely accepted solution is that it is due to
AGN feedback. In many models, massive galaxies merge to generate a
massive black hole surrounded by dense gas. The gas feeds both star
formation and an active nucleus (e.g. Sanders et al 1988). The power
of the AGN blows the gas away leaving a red, dead elliptical galaxy
(e.g. Springel, Di Matteo \& Hernquist 2005).

Studies of the colours of elliptical galaxies indicate that galaxies
move on a colour--magnitude diagram from the blue cloud of star
forming galaxies to the red cloud of dead ones. Interestingly most of
the hosts of AGN are found in the ``green valley'' between these two
extremes (Nandra et al 2007; Schawinski et al 2007). The rate at which
the galaxies have changed in colour can be deduced from post-starburst
signatures in their spectra, and appears to be a few 100 million
years, significantly faster than would be expected from passive
evolution, where stellar mass loss would accumulate and lead to late
star formation.

The observational details of this are however uncertain in local
galaxy bulges (Wild, Heckman \& Charlot 2010), with signs that the
black hole fuelling may lag behind the starburst (see Hopkins 2011 for
a model). Bell
et al (2011) studying massive galaxies from $0.6<z<2.2$ find
quiescence to correlate poorly with stellar mass. A common factor of a
quiescent galaxy is that it has a bulge, with presumably a central
black hole consistent with black hole feedback. A possibly important
uncertainty on distant AGN hosts is whether they are dusty or not
(e.g. Brammer et al 2009). Correcting for dust may remove most AGN
from the green valley altogether (Cardamone et al 2010).

\subsection{Mergers or secular evolution?}

Many theoretical models for quasar evolution are based on
galaxy-galaxy mergers being the trigger for gas infall onto a black
hole. (This is a convenient assumption since the merger rate is
predictable from the growth of large-scale dark matter structure.)
Although mergers must occur, the evidence for them triggering AGN is
weak at most redshifts. Searches for post-merger disruption signatures
often give a null result when the host galaxies of AGN are compared
with a control sample of field galaxies (e.g Cisternas et al 2011;
Schawinski et al 2011). Mergers are best seen as a trigger for distant
SubMillimetre Galaxies (SMG, Tacconi et al 2008; Engel et al 2010; Riechers
et al 2011) and some local Seyferts (Koss et al 2010). Secular
processes may nevertheless dominate gas inflow in massive bulges at
$z\sim2$ (Genzel et al 2009).

The evidence is building that between redshifts of 2 and the current
epoch, much of the evolution of AGN is secular (Kocevski et al 2011,
Orban de Xivry et al 2011). Further evidence for this emerges from a
study of the probability that a galaxy hosts an AGN. Aird et al (2011)
find this to be a power-law in Eddington rate, and largely independent
of mass (see also Kauffmann \& Heckman 2009 for a discussion of
Eddington ratios at low redshift and Alexander \& Hickox 2011 for a
review of black hole growth).
  
Secular evolution has  implications for the spin of  black
holes, which are then likely to be high (Berti \& Volonteri
2007). Hints that most accretion takes place onto spinning black
holes, with consequent high radiative efficiency, $\eta$, have emerged
from application of Soltan's (1982) argument relating the energy
density of quasar/AGN radiation to the local mean mass density in
massive black holes.
$${\cal E}(1+z) = \eta \rho_{\rm BH} c^2,$$
where $\cal E$ is the energy density in radiation from accretion, $z$
is the mean redshift at which the energy is radiated and $\rho_{\rm
  BH}$ is the mean smoothed-out density in black holes at the present
epoch.  The equation is independent of cosmological model and reflects
the fact that both the black hole mass and energy radiated remain and
scale together apart from the $(1+z)$ redshift factor that must be
applied to the radiation. Application of this formula to the X-ray
background or quasar counts etc, usually yield a value for $\eta$ of
0.1 or more (Fabian \& Iwasawa 1999; Elvis, Risaliti \& Zamorani 2002;
Marconi et al 2004; Raimundo \& Fabian 2009). This is higher than the
efficiency of a non-spinning black hole $\eta=0.057$ and consistent
with moderate to high spin.

Mergers may still be the trigger for the quasar peak at redshifts of
2--3. Whether a merger is wet or dry (gas rich or poor) can have a
significant effect on the final merger product, as can how and whether
the massive black holes of the merging galaxies scours the final
galaxy core or not (Kormendy et al 2010).  

The picture emerging from many observations of massive galaxies and
AGN is of radiative feedback being an important process when the AGN/quasar was
highly luminous and within about two orders of magnitude of the
Eddington limit. For massive galaxies, this highlights the redshift
range of the quasar peak. We now shift attention to low redshifts and
the most massive galaxies at the centres of clusters and groups. They
generally do not host luminous AGN or quasars. They do host the most
massive supermassive black holes, and are often active radio
sources. Feedback takes place here through the kinetic mode involving
jets acting on hot gas.

\section{The Kinetic Mode}

The more massive galaxies at the centres of groups and clusters are
often surrounded by gas with a radiative cooling time short enough
that a cooling flow should be taking place (Fabian 1994). The X-rays
we see indicate a large radiative loss and   mass cooling rates
of tens, hundreds or even thousands of $\Msunpyr$: 
$$\dot M={2\over 5}{{L\mu m}\over{kT}},$$
with a (factor roughly two) downward correction for gravitational
infall if the hot gas flows inward as a consequence of cooling. $\mu
m$ is the mean mass per particle of the gas of temperature $T$ and $L$
is the luminosity (mostly emitted in the X-ray band).

Some relevant gas properties of a small sample of objects are shown in
Fig.~2, ranging from the high luminosity cluster A\,1835 through the
X-ray brightest cluster in the Sky, A\,426 (the Perseus cluster), the
low-mass cluster A\,262 cluster to the Milky-Way mass elliptical
galaxy NGC\,720. All the clusters show a large central temperature
drop within the inner 100~kpc and all objects show a radiative cooling
time dropping below $10^9$~Gyr within the inner 10~kpc. An approximate
mass cooling rate, in the absence of an heat source, can be deduced by
dividing the gas mass within a chosen radius by the cooling time at
that radius. If a cooling flow is operating, the mass cooling rates
need to be worked out cumulatively including gravitational work done,
which will increase the rates by a factor of 1.5--2, depending on the
details of the profiles.

\begin{figure}
\begin{center}
\includegraphics[width=6in]{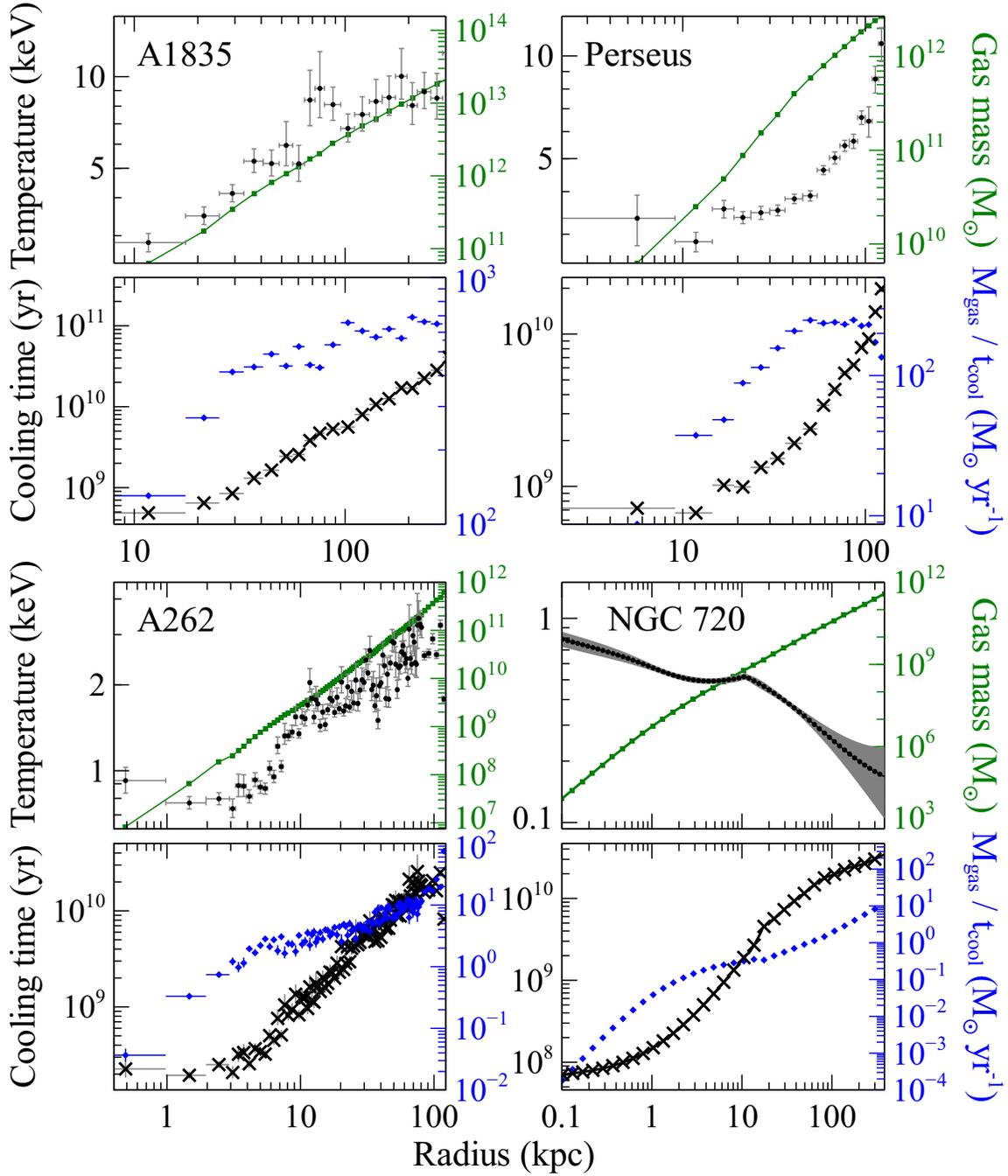} 
\caption{Gas temperature (+), cumulative gas mass (joined dots),
  $M_{\rm gas}(<r)$, radiative cooling time (x), $t_{\rm cool}(r)$,
  and mass cooling rate (-), $\dot M=M_{\rm gas}(<r)/t_{\rm cool}(r)$
  ), where $r$ is radius, for A\,1835 (Schmidt et al 2001; McNamara et
  al 2006), A\,426 (the Perseus cluster; Fabian et al 2006), A\,262
  (Blanton et al 2004; Sanders et al 2010a), and Milky Way mass
  elliptical galaxy NGC\,720 (Humphrey et al 2011), composite courtesy
  of J. Sanders. The temperatures shown here are deprojected values
  and assume single phase gas. Spatial and further spectral studies
  often show it to be multiphase near the centre.  }
   \label{fig1}
\end{center}
\end{figure}

The mass cooling rates are such that the clusters should be
significantly growing their stellar mass now, if radiative cooling is
uninhibited and the cooled gas forms stars. Observations do reveal
some star formation taking place, and A1835 may have the highest star
formation rate in a low redshift BCG ($\sim125\Msunpyr$; Egami et al
2006), but it does not equal the uninhibited mass cooling rate which
is $\sim 1000\Msunpyr$. Only if the initial mass function (IMF) of the
star formation process in these systems favoured low mass stars could
there be sufficient stars. The high pressure environment in a cluster
core, where the thermal pressure is about 1000 times that of the
interstellar medium of the Milky Way has been invoked as an
explanation for low mass stars, due to its effect on the Jeans mass
(Fabian et al 1982). Most observations of the IMF in a wide variety of
objects support a universal IMF which does not have most of its mass
in very low mass stars, which would be required here. Van Dokkum \&
Conroy (2010) do however find an IMF rich in low-mass stars in a small
sample of nearby elliptical galaxies (see also Cappellari et al 2012),
so the case may not yet be completely closed for low-mass stars
playing a role.

The centres of A\,1835 and the Perseus cluster do contain extensive
reservoirs of dusty atomic and molecular gas, the mass of which could
be the end result of a significant cooling flow, except that it would
not then be clear where the dust and molecules formed.  It is
generally considered that dust cannot form spontaneously in diffuse
cooled gas, which is presumably required first in order to then form
molecules. The X-ray rich environment in clusters does however mean
that the H$^-$ route may be open for molecule formation in cold
gas. Dust formation from cold molecular gas has been proposed in this
situation (Fabian, Johnstone \& Daines 1994) but no detailed
calculation has been attempted. The situation may change now that
significant amounts of dust and molecules have been found in some very
young supernova remnants, such as SN1987A (Matsuura et al 2011) and the
Crab (Loh, Baldwin \& Ferland et al 2011).

XMM-Newton Reflection Grating Spectrometer observations provided
crucial information against a simple cooling flow model in that they
failed to show the strong lines expected from FeXVII as the gas cooled
below 0.7~keV (Peterson et al 2001; Tamura et al 2001).  Detailed fits
(Peterson et al 2003) indicated that there was much less gas 
below one third of the outer cluster gas temperature than would be
expected in a steady cooling flow. Either something was heating the
gas, or the gas was somehow disappearing. As will be discussed below, both
of these options are probably involved. 

The likely heat source is the AGN in the BCG at the centre of the cool
core. Almost all have an active radio source (Burns 1990, Sun
2010). Heating by the central AGN was suggested early by Pedlar et al
(1990), Baum \& O'Dea (1991), Tabor \& Binney (1994) and Tucker \&
David (1997). Later work shows that the correlation between radio
power and cooling luminosity (a measure of the rate at which gas cools
within a fiducial cooling radius where $t_{\rm cool}$ is 7~Gyr) is
poor (Voigt \& Fabian 2004) and the jets and thus the kinetic power
either has to be highly sporadic or extremely radiatively inefficient
(see later).

The general consensus now is that the massive black hole at the centre of
the galaxy is feeding energy back into its surroundings at a rate
balancing the loss of energy through cooling (for reviews see Peterson
\& Fabian 2006, McNamara \& Nulsen 2007, Cattaneo et al 2009).

Several steps in this feedback process are clearly seen in X-ray and
radio observations. Much of the action is spatially resolved and the
gas optically thin.  The accretion flow onto the black hole generates
powerful jets which inflate bubbles of relativistic plasma either side
of the nucleus. The bubbles are buoyant in the intracluster or
intragroup medium, separating and rising as a new bubble forms
(Churazov et al 2000; McNamara et al 2000). A study of the brightest
55 clusters (Fig.~3; Dunn \& Fabian 2006, 2008) originally showed that
over 70\% of those clusters where the cooling time is less than 3~Gyr,
therefore needing heat, have bubbles; the remaining 30\% have a
central radio source. This implies that the duty cycle of the bubbling
is at least $70\%$. Updating that work now indicates (Fig.~3) that the
bubble fraction is $19/20$ objects (the odd one out -- the Ophiuchus
cluster -- is undergoing a merger into the core). When projection
effects are considered, since bubbles along the line of sight will be
difficult to distinguish, the corresponding bubbling fraction is
$>95$\%. Note that objects cannot shift to the left in this diagram on
timescales less than $t_{\rm cool}$; they can shift to the right on a
shorter timescale but this must be rare or there would not be a peak
at low $t_{\rm cool}$.  {\em The jet bubbling process is not therefore
  very episodic, but is more or less continuous.}

\begin{figure}
\begin{center}
\includegraphics[width=4.in]{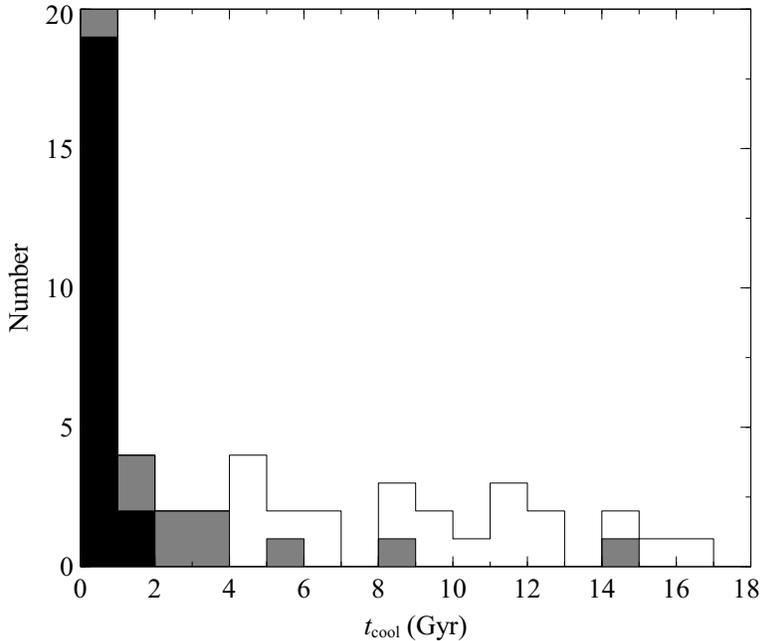} 
\caption{ Histogram of cooling times in the B55 cluster sample
  (updated from Dunn \& Fabian 2006). Black indicates bubbles seen and
  grey that there is a central radio source. The plot has been updated
  using Chandra cooling times and later imaging, which has revealed
  more bubbles. The cooling time distribution is similar to that
  reported by Mittal et al (2009) which uses an overlapping, soft
  X-ray selected, sample. The only source in the first bin lacking
  bubbles is the Ophiuchus cluster, which appears to be undergoing a
  core merger (Million et al 2010). }
   \label{fig1}
\end{center}
\end{figure}

The kinetic power in the jets can be estimated from the product of the
volume of the bubbles (Fig.~4), and the surrounding pressure (obtained
from the density and temperature of the thermal gas), divided by the
buoyancy time (which depends on the gravitational potential). The
power is high and only weakly correlated with radio power (the
radiative efficiency of many jets is very low at between
$10^{-2}-10^{-4}$). The power is usually in good agreement with the
energy loss by X-radiation from the short-cooling-time region (Fig.~5,
McNamara \& Nulsen 2007, Rafferty et al 2006, 2008).  The overall
energetics of the feedback process are therefore not an issue.

\begin{figure}
\begin{center}
 \includegraphics[width=6in]{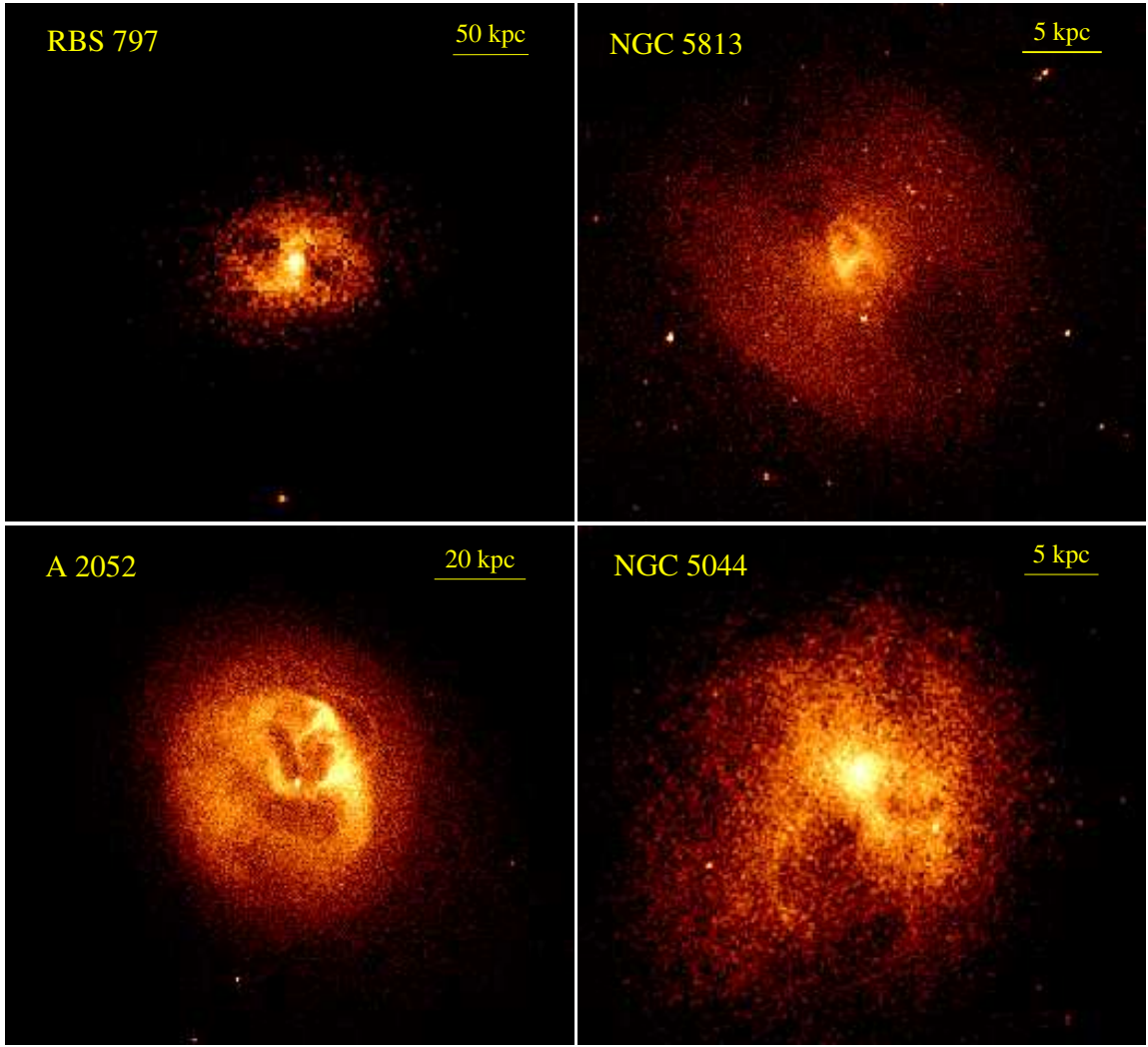}
 \vspace*{0 cm}
 \caption{Chandra X-ray images showing the dramatic interaction of the
   central AGN on the surrounding gas over a range of scales. Top
   left: massive cluster RBS\,797 at $z=0.354$ (Cavagnolo et al 2011),
   nearby central group ellipical galaxy NGC\,5813 at $z=0.006$
   (Randall et al 2011), Lower left: rich cluster A\,2052 at $z=0.035$
   (Blanton et al 2011) and NGC\,5044 group at $z=0.0093$ (David et al
   2011). Note that the bubbles in RBS\,797 have volumes about 1000
   times larger than those of the inner bubbles of NGC\,5813. Another
   larger pair of bubbles occur in NGC\,5813, making 3 pairs in all.}
   \label{fig1}
\end{center}
\end{figure}

\begin{figure}
\begin{center}
 \includegraphics[width=5in]{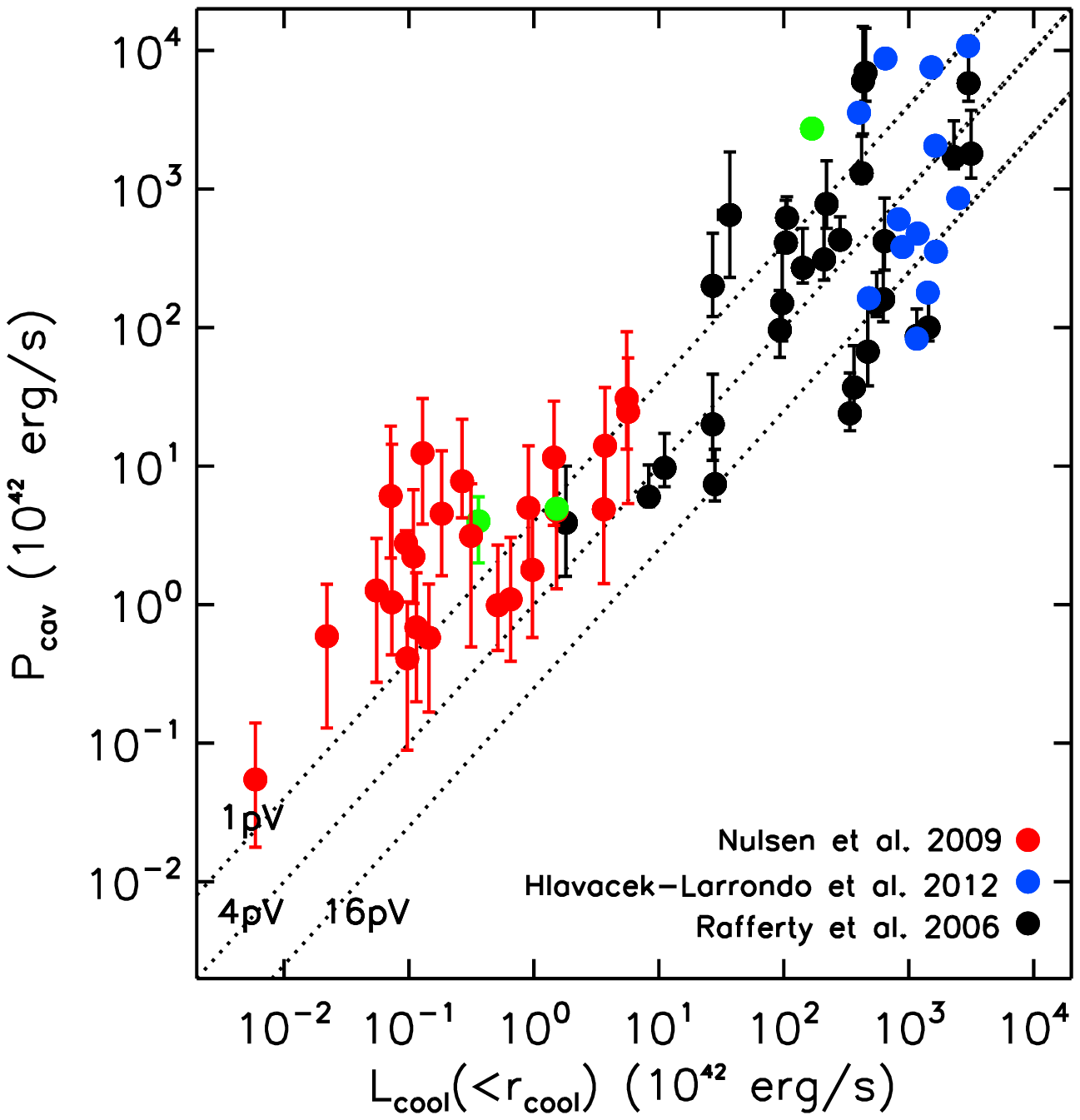} 
 \caption{Power inferred from cavities/bubbles plotted against
   luminosity within the cooling region (where the radiative cooling
   time is less than 7~Gyr), courtesy of J. Hlavacek-Larrondo. The
   objects range from luminous clusters, through groups, to elliptical
   galaxies.  }
   \label{fig1}
\end{center}
\end{figure}

\subsubsection{Bubbles}

The bubbles, or cavities, commonly seen in deep Chandra images of cool
core clusters are blown and powered by jets from the central black
hole. A Fanaroff-Riley (1974) type I radio source usually coincides neatly
with the cavity.  The idea that a radio source could blow bubbles in the
ICM dates back to a paper by Gull \& Northover (1973). The NorthWest
outer bubble in the Perseus cluster, called a ``ghost'' bubble because
of a lack of high-frequency radio synchrotron emission due to radio
spectral ageing, was seen in Einstein images (Branduardi-Raymont et al
1981, Fabian et al 1981) but not recognised as such back then. The
first clear image of bubbles was made using the ROSAT HRI of the
Perseus cluster core by Boehringer et al (1993), followed by many
observations with Chandra after its launch in 1999.

As already mentioned, most cool cores in the X-ray brightest clusters
with central radiative cooling times less than 3 Gyr, have clear
bubbles seen by Chandra. Allowing for projection effects, the real
fraction is higher. The innermost bubbles are usually fairly spherical
and in the best-studied case (the Perseus cluster) are surrounded by a
thick high-pressure region fronted by a weak shock (Fig.~6). The
thermal energy within that region corresponds to 3.7 times that of a
surrounding region of similar volume to the bubble (Graham, Fabian \&
Sanders 2008), indicating that bubbles can transfer almost $4PV$,
which is the internal energy expected from a relativistic fluid
($\gamma=4/3$). The mechanical power of a new bubble is often assumed
to be $4PV/t_{\rm age}$, where $t_{\rm age}$ is the risetime of the
bubble (Churazov et al 2002).

Bubbles rise buoyantly in the surrounding hot atmosphere, turning into
ghost bubbles as they become undetectable in (high frequency) radio
observations. There are far fewer of these outer bubbles known, but
they can appear at substantial radii and are then larger than expected
(Diehl, Fryer \& Rafferty 2008: note the sequence of bubbles in
NGC\,5813 shown in Fig.~4). This could be due to their rate of rise
being a function of their size, i.e. larger ones move slower catching
up smaller faster ones, or vice versa (Fabian et al 2011). Observed
bubbles appear to be fairly stable to breakup, contrary to what is
seen in many simulations. Air bubbles in water can be quite large and
stable. The action of blowing a bubble means that the surface is never
static with respect to its surroundings so it is not Rayleigh-Taylor
unstable. The growth time of large scale Kelvin-Helmholtz
instabilities is similar to the flow time so this need not be a
problem either. Provided that something like viscosity (Reynolds et al
2005) or magnetic draping (Lyutikov 2006) can damp the smallest
perturbations, then there is no immediate reason for them to break up.
   
\subsubsection{Energy flow and dissipation}
Having enough energy available does not indicate how the energy flows
and is dissipated. The coolest material is found next to the heat
source, which is not what would be expected of a heat source in a
room, say. The jets from the central black hole are intrinsically
anisotropic. The jet/bubble axis is not apparent in temperature maps
of cluster cores, so heating is presumably much more isotropic. As
discussed below, linewidths show that the level of turbulence in the
gas is relatively low (energy in turbulence is less than 10 per cent
of the thermal energy) and steep abundance gradients show that there
is no large scale, violent, mixing taking place. The bubbling process
is relatively gentle and continuous, perhaps rather like a dripping
tap (or a fishtank aerator).

In the case of the Perseus cluster, which is the X-ray brightest in
the Sky, Chandra imaging shows concentric ripples which we interpret
as sound waves generated by the expansion of the central pressure
peaks associated with the repetitive blowing of bubbles (Fig.~5,
Fabian et al 2003, 2006). The energy flux in the sound waves is
comparable to that required to offset cooling, showing that this is
the likely way in which heat is distributed in a quasi-spherical
manner. Similar sound waves, or weak shocks, are also seen in several
of the very brightest clusters in the Sky, e.g., the Virgo (Fig.~8,
Forman et al 2005), Centaurus (Sanders et al 2008) and A2052 (Blanton
et al 2011) clusters and in simulations (Ruszkowski, Br\"uggen \&
Begelman 2004; Sijacki \& Springel 2006). The amplitude of the ripples
is less than 10 per cent, so they will be very difficult or impossible
to see in less bright clusters (Graham et al 2008b).

\begin{figure}
\begin{center}
\includegraphics[width=5.1in]{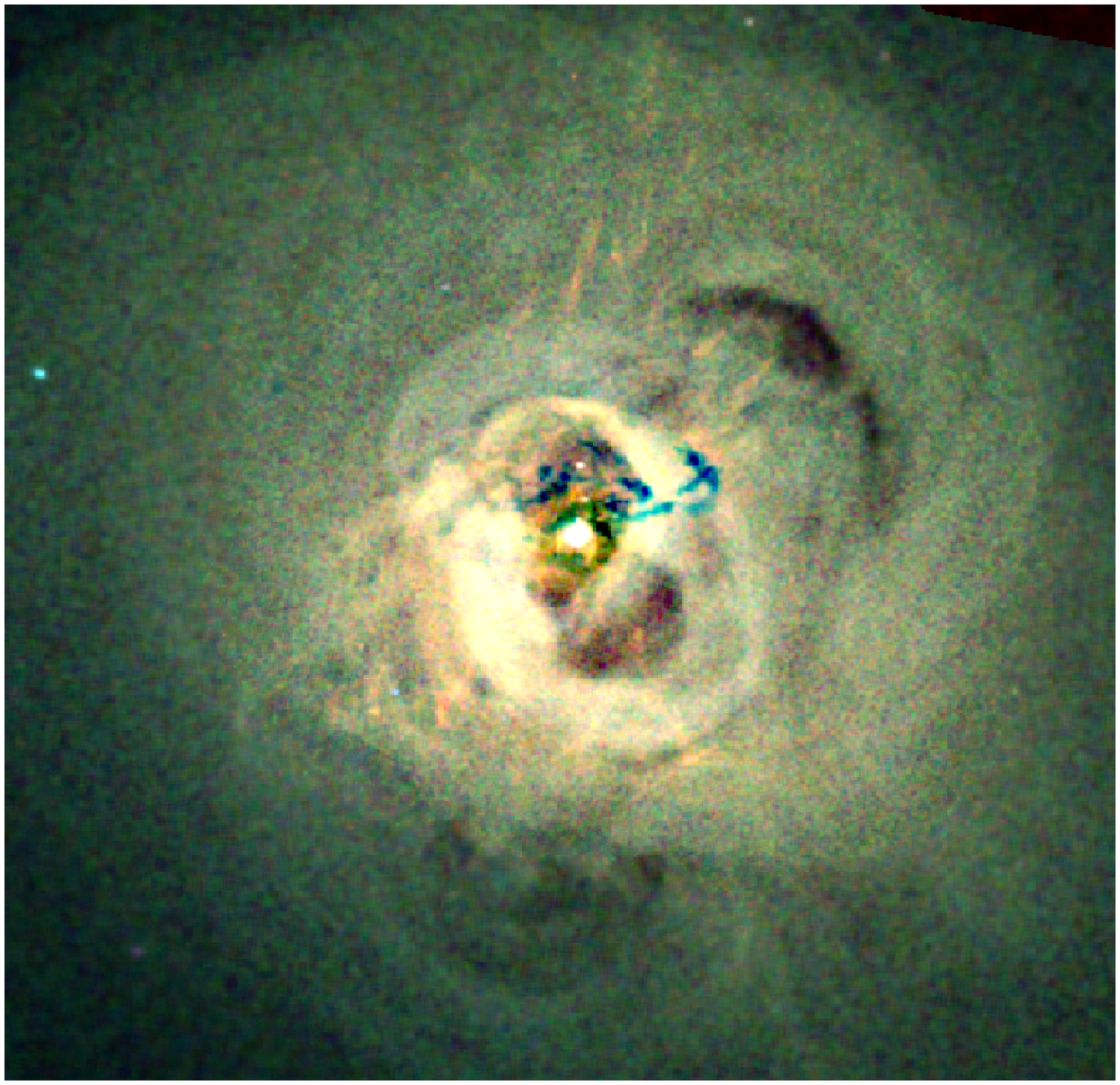}
 \includegraphics[width=2.6in]{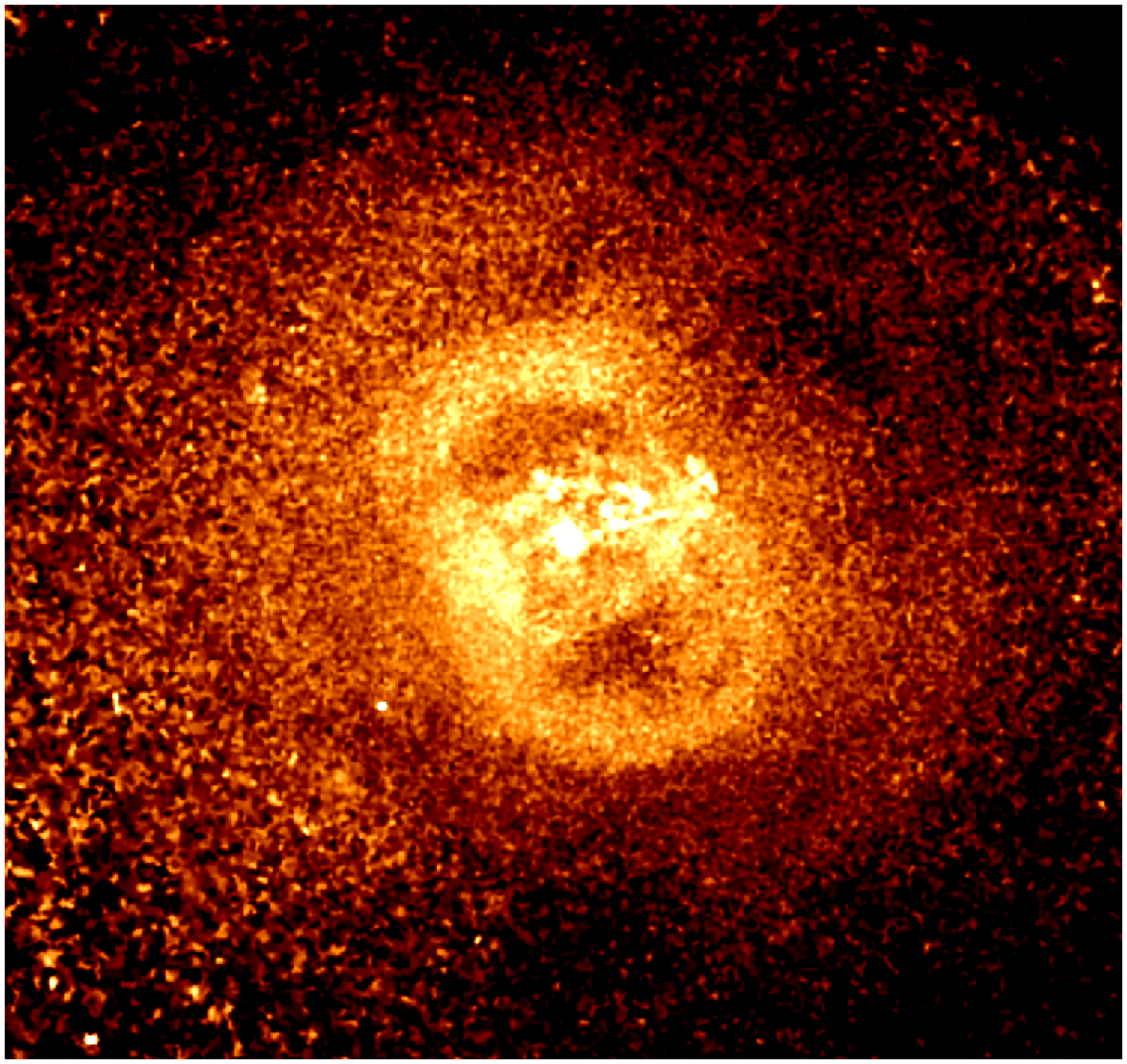}
 \includegraphics[width=2.45in]{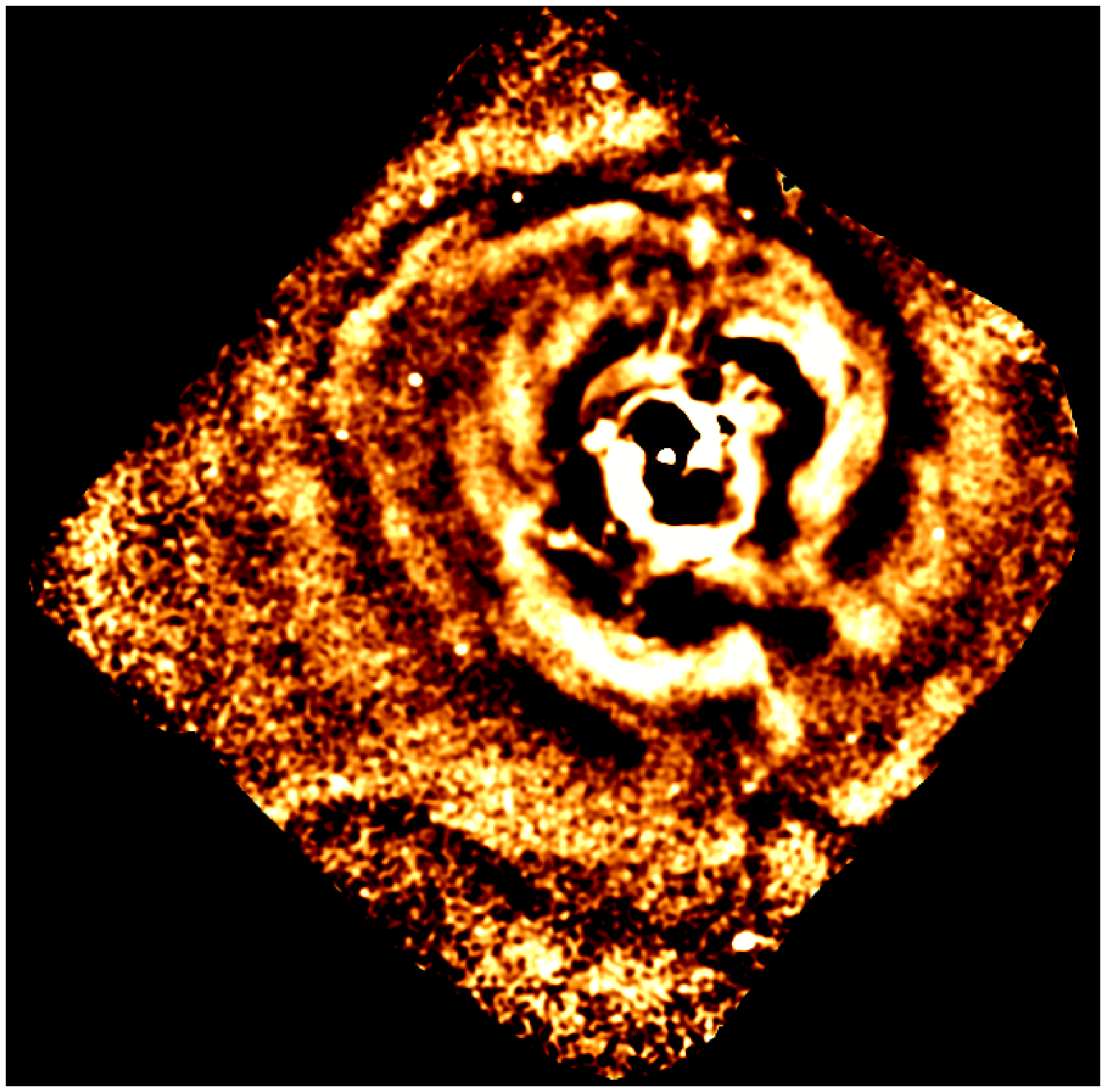}
 \caption{Top: Chandra X-ray image of the Perseus cluster
   core. Red--Green--Blue depicts soft to hard X-rays. The blue
   features near the centre are due to absorption by the infalling
   high velocity system, a galaxy which must lie at least 100~kpc
   closer to us in the cluster (otherwise the absorption would be
   filled in with cluster emission). Note the clear inner and outer
   bubble pairs as well as the weak shock to the North East of the
   inner Northern bubble. Lower Left: Pressure map derived from
   Chandra imaging X-ray spectroscopy of the Perseus cluster. Note the
   thick high pressure regions containing almost $4PV$ of energy
   surrounding each inner bubble, where $V$ is the volume of the
   radio-plasma filled interior (Fabian et al 2006). Lower Right:
   unsharp-masked image showing the pressure ripples or sound waves. }
   \label{fig1}
\end{center}
\end{figure}

Weak shocks are poor at dissipating energy, so the heating of the gas
must depend on the gas viscosity.  The Spitzer--Braginsky viscosity
for an ionized gas yield a dissipation length on the order of 100~kpc,
so in the required range (Fabian et al 2005). The intracluster gas is
however magnetised (as inferred from Faraday Rotation measurements,
see e.g. Taylor et al 2007), in which case the value of the viscosity
is not clear (see Kunz et al 2010; Parrish et al 2012; Choi \&
Stone 2012). It is important to realise that for transport purposes,
much of the relevant intracluster gas cannot be classified as either
collisional or collisionless but is somewhere in between.

\subsubsection{Abundance Gradients}

The inner parts of most cool cores show higher metal abundance that
the bulk of the intracluster gas. The metallicity can reach 2 to 3
times Solar values at the peak. This is considered to be due to
pollution by stars and supernovae of the BCG. The spatial breadth of
the high abundance region is broader than expected from a static
atmospher, probably due to some turbulence and motions caused by the
central AGN. The existence of the peaks argues against these motions
being too disruptive, or of the feedback having been too violent in
the past. Constraints on the level of turbulence have been obtained
from comparison of the abundance peaks with the underlying BCG light
profile by Rebusco et al (2006) and Graham et al (2007).

Rising bubbles appear to have dragged some of the lower entropy,
metal-enriched gas outward in some cores (Werner et al 2010, 2011;
Simionescu et al 2009).

\subsection{Heating/Cooling balance -- Maintenance mode feedback}

Let us now consider how close the apparent heating/cooling balance is
and how it has been established and maintained. The lack of high
star formation rates suggests that cooling does not exceed heating by
ten per cent or so. The presence of central abundance gradients and
pronounced temperature drops indicates that heating does not generally
exceed cooling by much either. This represents a relatively close balance
which needs to continue over tens to hundreds of bubbling cycles (each
of a 10--50~Myr or so, depending on power).

A simple 1D feedback cycle seems at first sight possible. If too much
gas starts to cool then the accretion rate should increase making the
heating rate go up and vice versa.  However the lengthscales involved
range over a factor of $10^9$ or more and the timescales involved over
the whole cooling flow region are long, up to and beyond a Gyr, and
down to a Myr at the accretion radius. This means that feedback would
be delayed or at least that there could be serious hysteresis. Angular
momentum could easily prevent gas reaching anywhere near the black
hole. These issues have been discussed most recently by Pizzalato \&
Soker (2010) and Narayan \& Fabian (2011).

\subsubsection{Bondi accretion}

Bondi accretion (Bondi 1957) could be relevant here, since it relies
on a point mass embedded in a static medium.  The gas accreting onto
the central black hole passes at least through about 5 orders of
magnitude in radius from where the gravitational field of the central
mass begins to dominate gas motions, i.e. the Bondi radius, to the
centre.  The flow itself originates further out. It is
implausible that angular momentum can be ignored. One approach is to
assume that the accretion flow is viscous all the way so angular
momentum is efficiently transported outward at all radii. The whole
inner region of radii a few 100~kpc, from beyond the Bondi radius to
the innermost regions where the flow becomes supersonic, could
resemble a giant Advection Dominated Accretion Flow (ADAF) (Narayan \&
Fabian 2011). This does at least allow for easy passage of the gas
without it becoming choked by angular momentum.

An ADAF is radiatively inefficient since it transports (advects) the energy
released in with the gas. Several percent (Allen et al 2006) or more
of the power in the accretion flow  must  be released into a
(radiatively inefficient jet) close to the black hole for this
hypothesis to be viable. There are indications from the observed behaviour of
Galactic black hole binaries (Section 4.2) that this can happen,
although the details are poorly understood. 

Good observations of the inner regions of the flow onto a massive
black hole in a BCG are of M87 at the centre of the Virgo cluster. The
Bondi radius there is 100--200~pc, corresponding to 1--2 arcsec (Di
Matteo et al 2003). There is no evidence for a radiatively-efficient
accretion disc extending in to the black hole. The nearside jet is
well seen and resolved down to very close to the black hole (it is
accelerated within $100 r_{\rm g}$;  Hada et al 2011).

Simple Bondi accretion appears not to work in some objects. For
example, the Bondi radius of the nearby, $\sim10\Mpc$, $\sim
10^9\Msun$ black hole in NGC\,3115 is at $\sim 5$~arcsec so well
resolved in X-rays by Chandra. The nucleus X-ray luminosity of
$<10^{38}\ergps$ is $<10^{-7}$ times that predicted by Bondi accretion
from the observed surrounding hot halo (Wong et al 2010). Energy
feedback within the Bondi radius may be responsible.  

Bondi accretion has sometimes been considered insufficient for
powerful objects (Rafferty et al 2006; Hardcastle, Evans \& Croston
2007; McNamara, Rohanizadegan \& Nulsen 2011). The issue is considered
again in Section 3.4.
 
\subsubsection{Temperature structure}

As mentioned earlier, studies with Chandra and XMM suggested that the
coolest X-ray detectable gas was at a temperature one-third of the
cluster virial temperature (Peterson et al 2003). However careful work
on the best data from the brightest objects shows that the temperature
range extends to a factor of at least ten (Fig.~7 left, Sanders et al
2008).  There is however less and less gas found at lower temperatures
compared with say a steady cooling flow (Fig.~7 right). Interpreted
from a cooling flow perspective, the gas with the shorter, and
particularly the shortest, cooling times appears to be missing.

\begin{figure}
\begin{center}
 \includegraphics[width=3.3in]{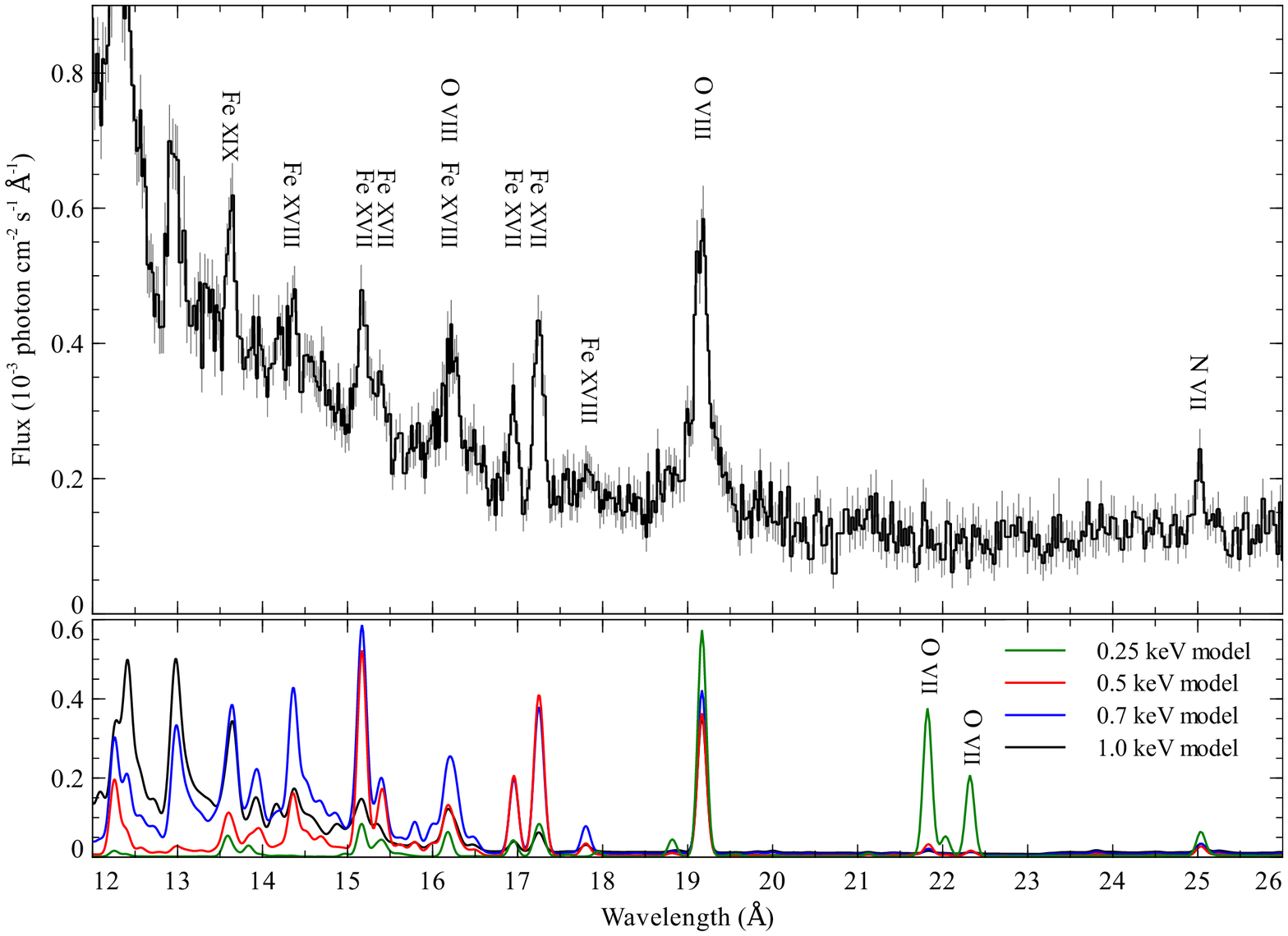} 
\hspace{0.5cm}
\includegraphics[width=2.4in]{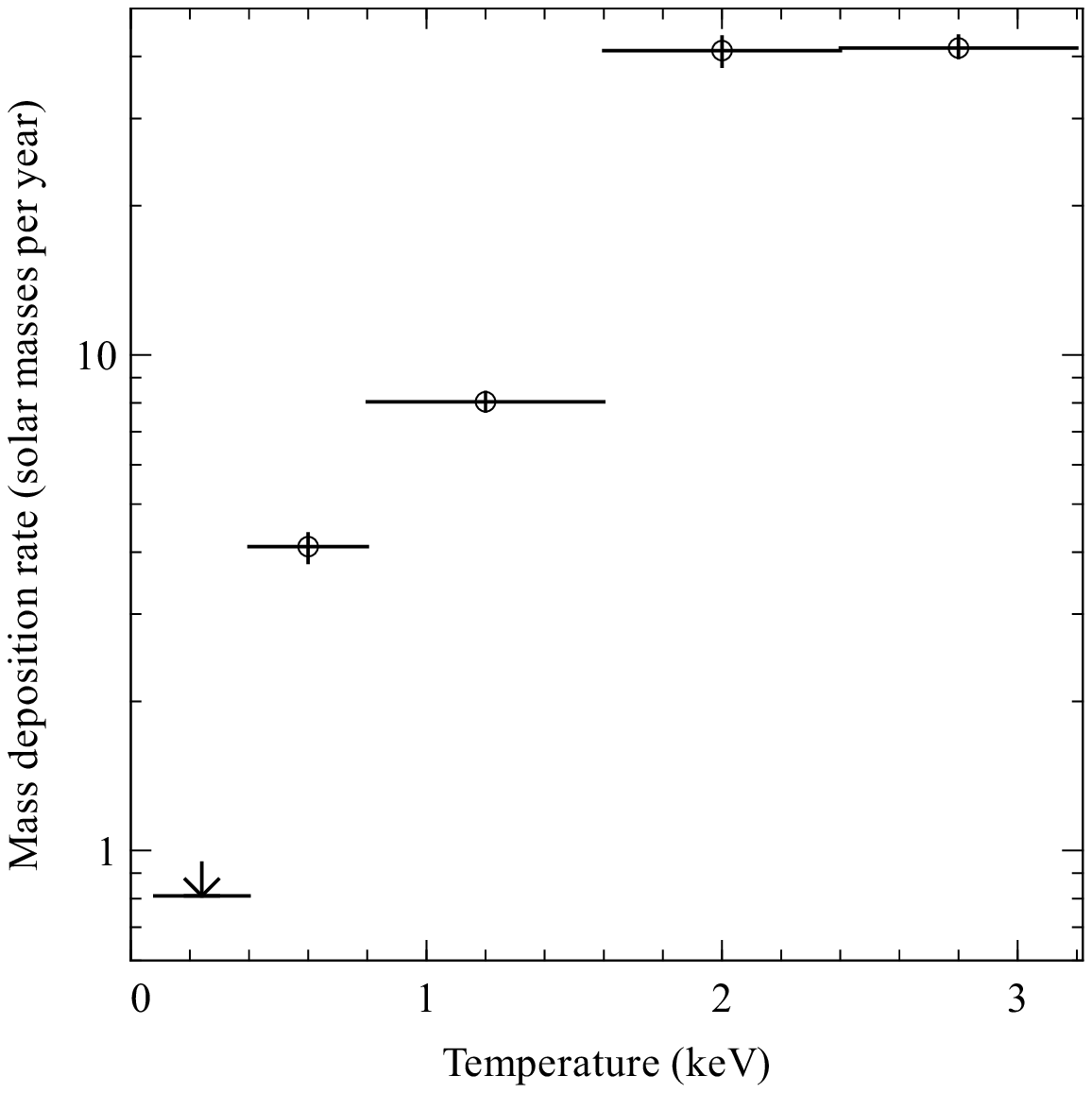}
\caption{Left: XMM-RGS spectrum of the centre of the nearby Centaurus
  cluster showing strong FeXVII at 15 and 17~A and OVIII at 19~A lines
  but no OVII at 22~A (Sanders et al 2008). Right: Strength of soft
  X-ray emission in the Centaurus cluster core, interpretated as due
  to a mass cooling rate of the gas. Note that little gas seems to
  cool below 0.5~keV (OVII is emitted by gas at $\sim 0.3\keV)$. There
  is no continuous radiative cooling flow here, although some
  non-radiative cooling may occur by hot and cold gas mixing.  }
   \label{fig1}
\end{center}
\end{figure}

X-ray images from Chandra and moderate resolution spectra from the
XMM-Newton Reflection Grating Spectrometer (RGS) show X-ray cool gas
ranging from 5 to 0.5~keV in the nearby Centaurus cluster (Sanders et
al 2008).  The coolest gas has a radiative cooling time of only
10~Myr, yet the spectra show no sign of any lower gas temperature gas
(where detectable OVII emission is expected). In this object the
heating/cooling balance looks to hold to a few per cent. How
the 0.5~keV gas is prevented from cooling is not obvious. The X-ray
images show that it is clumpy so the question arises as to how it is
targetted for heating without its immediate surroundings being
overheated. A similar picture emerges from several other clusters with
excellent data.

One solution is that the tight balance is only apparent. If the jets
become too energetic then their intrinsic anisotropy dominates and
they can they push through the whole cooling region (e.g. Cyg A or
MS0735.6+7421; McNamara et al 2009, Fig.~8), depositing their energy
much further out. The problem here is that no low power Cygnus A
analogues (Fanaroff-Riley type II edge-brightened sources) are seen in
local BCGs.  If cooling dominates then it can feed the reservoir of
cold gas seen in many objects, as well as star formation. The BCG at
the centre of A1835 at $z=0.25$ is an extreme example with over
$\sim125\Msunpyr$ of massive star formation. It is
within a factor of two of the highest star formation rate of any
galaxy at low redshift (Arp 220 has a rate of about
$200\Msunpyr$). (Without heating, the central intracluster gas in
A1835 would be cooling at over $1000\Msunpyr$, so a balance remains,
but not a very tight one.)

\begin{figure}
\begin{center}
 \includegraphics[width=2.5in]{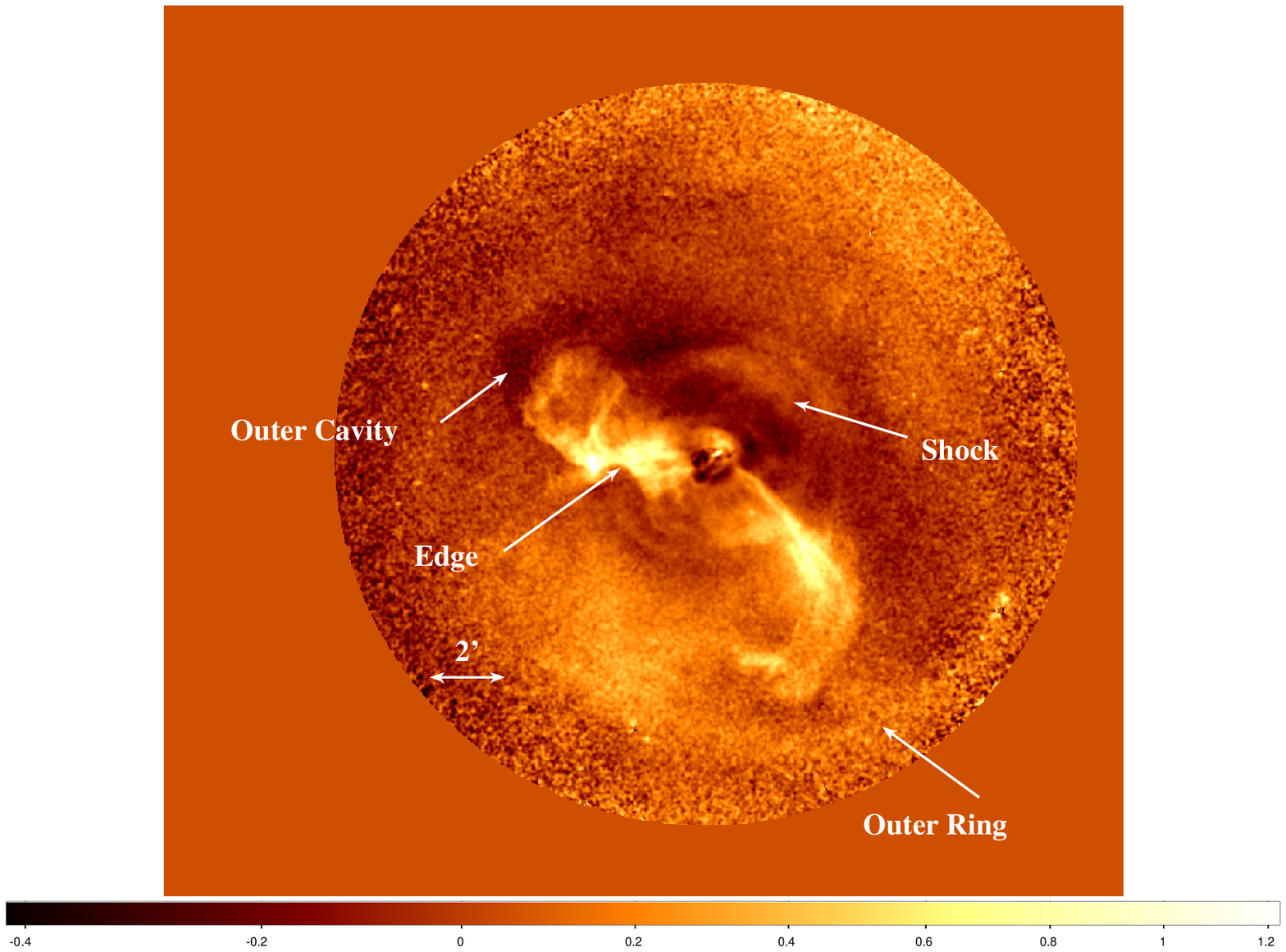}
\hspace{0.3cm}
\includegraphics[width=2.5in]{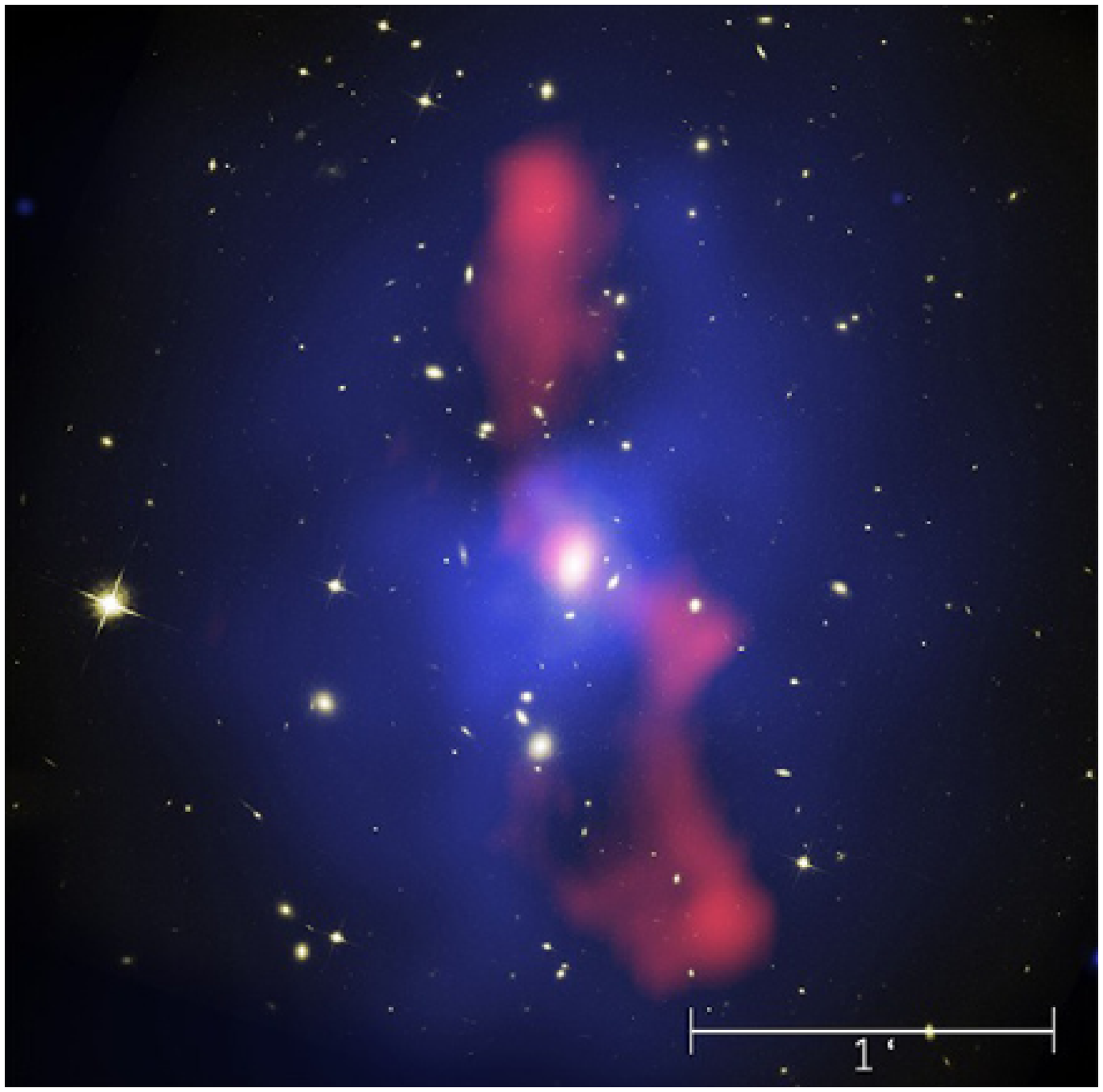}
\vspace*{0.5 cm}
\caption{Left: The Arms and weak shocks produced by the jets of M87
  (Forman et al 2007). Right: The gigantic interaction of the radio
  lobes and intracluster gas of MS0735.6 (McNamara et al 2009). The
  figure shows the inner 700~kpc of the cluster, extending well beyond
  its cool core.  }
   \label{fig1}
\end{center}
\end{figure}

\subsection{Cool, Cold Gas and Star Formation}

Many BCGs in {\it cool core} clusters (the ones with the short
radiative cooling times) have extensive optical emission-line
nebulosities (e.g. Crawford et al 1999). Indeed it is only those
clusters with central cooling times below 2~Gyr which have filamentary
nebulosities (Hu et al 85, Heckman et al 1989, Cavagnolo et al
2010). The optical spectrum is predominantly of low ionization gas and
is quite unlike any Galactic PhotoDissociation Region (PDR) such as
the Orion Nebula (Ferland et al 2009). The emission can extend for
tens of kpc around the BCG -- in the famous case of NGC\,1275 the
largest diameter of the nebulosity exceeds 80~kpc.

The bulk of the cold gas is molecular as shown by CO (Edge 2001;
Salom\'e \& Combes 2003) and H$_2$ emission (Jaffe et al 2001; Edge et
al 2002; Johnstone et al 2007).  NGC\,1275 at the centre of the
Perseus cluster is a spectacular example (Fig. 9, Fabian et al 2008)
with filaments composed of about $5\times 10^{10}\Msun$ of H$_2$
(Salom\'e et al 2006). Molecular emission is clearly resolved out to
25~kpc and beyond (Hatch et al 2005), showing excellent agreement
between the atomic and warm molecular structure (Lim et al 2012) as
well as the cold CO structures (Salome et al 2011).  The mass of the
molecular gas is comparable to the mass of all other gas, hot and
cold, within the central 10~kpc.  Star formation happens sporadically
in that galaxy with $\sim 20\Msunpyr$ occurring over the past
$10^8\yr$ in the South-Eastern blue loop (Canning et al 2010).

\begin{figure}
\begin{center}
 \includegraphics[width=2.8in]{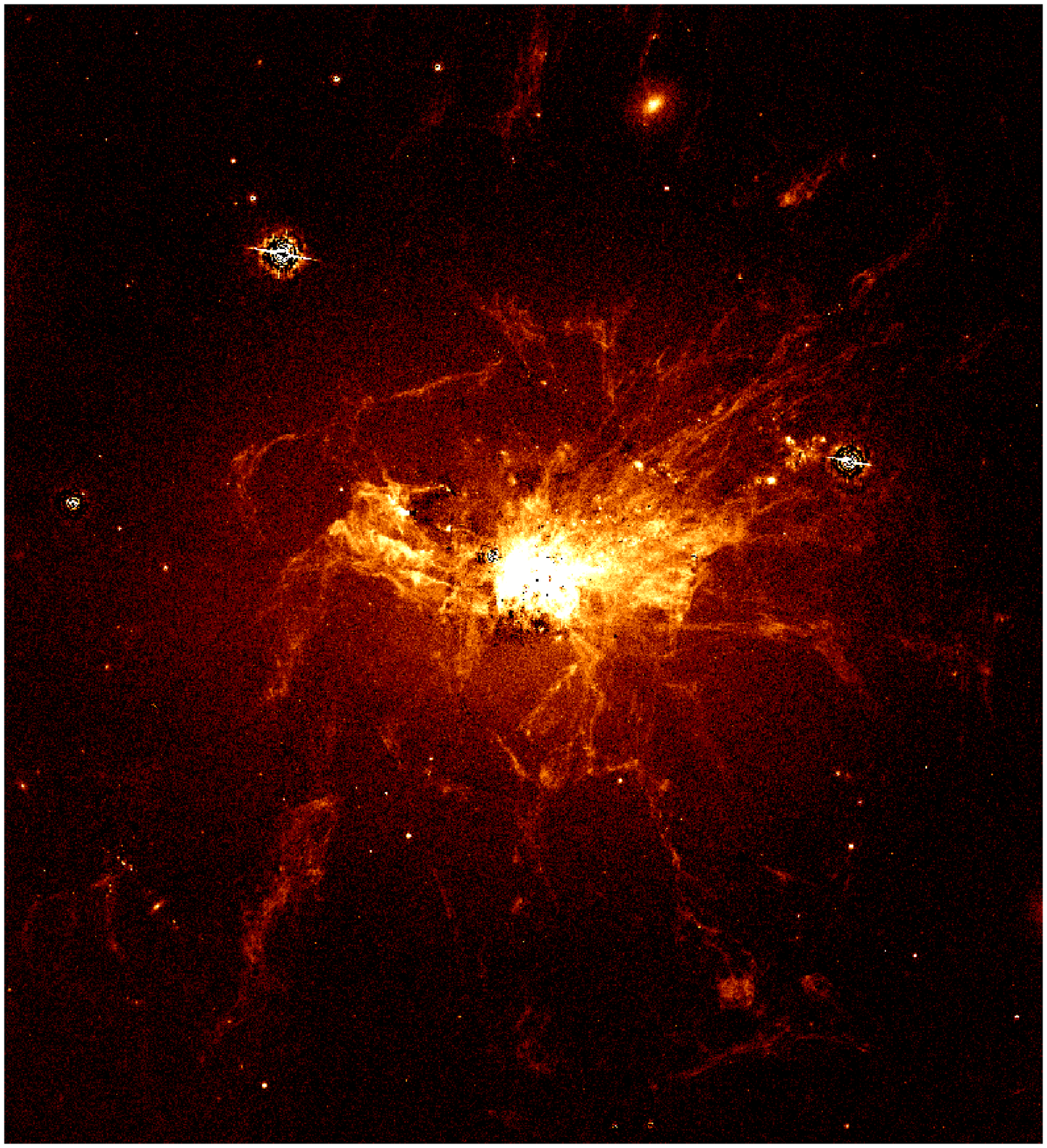}
\hspace{0.cm}
\includegraphics[width=3.1in]{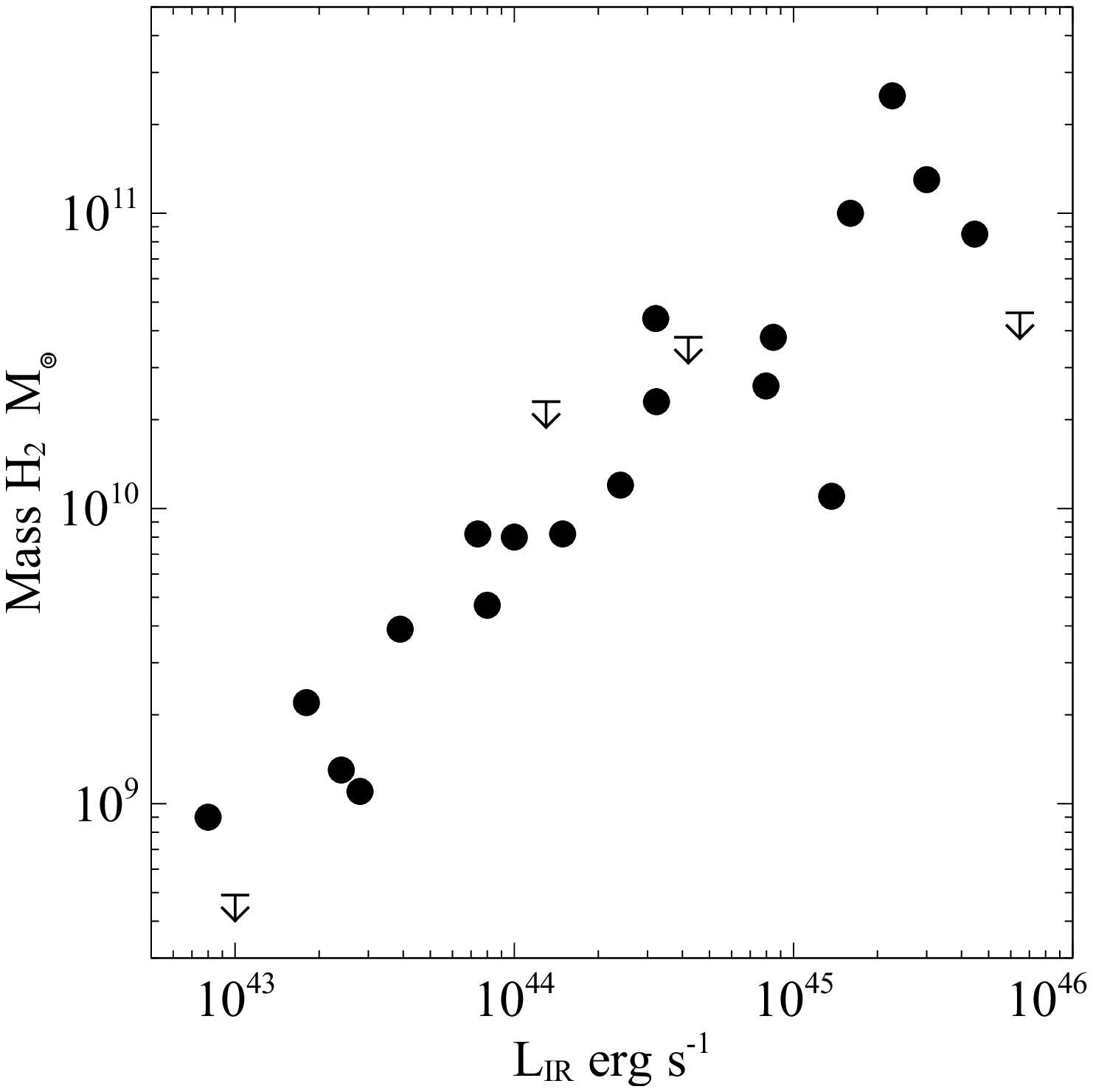}
\vspace*{0 cm}
\caption{Left: HST image of the filaments around NGC\,1275 in the
   Perseus cluster (Fabian et al 2008). Individual filaments are
   resolved at $\sim70\pc$ wide;  some are straight over 6~kpc
   lengths.  Right: Mass of cold H$_2$ reservoir, inferred from CO
   observations, in a sample of cool core BCGs compared with Spitzer
   IR luminosity (adapted from O'Dea et al 2008).  }
   \label{fig1}
\end{center}
\end{figure}

Dust is seen in many objects in the form of dust lanes and infrared
emission, with Spitzer observations revealing high IR luminosities
(Egami et al 2006, O'Dea et al 2008, Edge et al 2010,  Fig. 5). The dust
is presumably injected by stars into the central cold gas reservoir
(Voit \& Donahue 2011).

It is likely that many of the filaments have been dragged out from
near the centre of the cluster, within the BCG (Hatch et al
2006). This is supported by the horseshoe-shaped filament in the
Perseus cluster to the NorthWest of the BCG, NGC\,1275, lying behind
an outer bubble which is presumably rising buoyantly outward. The
smooth unruffled shape of the horseshoe resembles the streamlines
behind rising gas bubbles in water. This implies that the intracluster
medium has low turbulence on these scales and thus that the viscosity
is high (Fabian et al 2003).

Much of the total IR and UV luminosity of BCGs in luminous cool core
clusters is due to vigorous star formation in the BCG (C. O'Dea et al
2008; K. O'Dea et al 2010), presumably fuelled by a residual cooling
flow. Some of the IR luminosity however could be due to the coolest
X-ray emitting clumps, at 0.5--3~keV, mixing in with the cold gas and
thereby cooling non-radiatively (Fabian et al 2002; Soker, Blanton \&
Sarazin 2004). There is more than enough infrared emission in most
objects to account for a significant mass cooling rate.  The outer
filaments in NGC\,1275, generally show no embedded young stars and are
far from the central nucleus. The optical spectrum of the filamentary
gas here indicates low ionization, with strong Balmer lines, [OII],
OI, NI, [NII] and [SII], but weak [OIII] yet high
Ne[III]. Photoionization modelling fails to reproduce this spectrum if
O stars are assumed. It requires a much hotter incident spectrum with
a 150 thousand K blackbody spectrum being suitable (Crawford \& Fabian
1993). There is no such source of photons available, so it has been
concluded that the filaments cannot be excited by photons.

\subsubsection{Heating of cold gas by energetic particles}
An alternative which has been studied by Ferland et al (2008, 2009) is
that cosmic rays are the source of ionization and excitation. If they
penetrate cold gas then they collide with that gas and leave trails of
30--40~eV secondary particles which excite a spectrum resembling the
observed one. If the cold gas has denser molecular phases then the
H$_2$ and CO emission is also accounted for. The required energy
density in cosmic rays is high. Similar secondary particles are
however also produced by the incidence of keV particles, such as
constitute the surrounding hot gas.

This has lead to a variant of the cosmic-ray model in which the
filaments are powered by interpenetration of the hot gas (Fabian et al
2011). An interesting aspect of this model is that the incident flux
of energy onto a filament in the Perseus cluster from the surrounding
gas particles is only a few times higher than the total flux emitted
by the filament. This makes it energetically
feasible. Interpenetration does require that the magnetic fields which
must help support the filaments in the hot gas (Fabian et al 2008) and
give it integrity do not prevent the particles entering the
gas. Reconnection diffusion (Lazarian et al 2011) provides a mechanism
by which this can proceed. 

Since particles are entering a filament, its mass is slowly
increasing. The flux of particles is proportional to the
radation flux from the filament, which enables an estimate to be made
($\dot M\sim 70 L_{42} T^{-1}_7\Msunpyr$, where $L_{42}$ is the
H$\alpha$ luminosity in units of $10^{42}\ergps$ and $T_7$ is the
surrounding gas temperature in units of $10^7\K$). The filamentary
system in the Perseus cluster may thus be growing at $50-100\Msunpyr$,
which probably exceeds the current star formation rate. The mass of the
filaments doubles in about a Gyr.

Particle heating and other models are now being tested over a wide
range of spectra and imaging of BCG filaments, from the UV (Oonk et al
2011) including possible OVI emission (Oegerle et al 2001; Bregman et
al 2006), to optical (McDonald, Veilleux \& Rupke 2011; Canning et al
2011, Edwards et al 2009), near infrared (Oonk et al 2010), mid
infrared (Spitzer: Johnstone et al 2007; Donahue et al 2011) and far-infared
(Spitzer: Egami et al 2006; Herschel: Edge et al 2010; Mittal et al
2011) bands.  CO spectra and detection of HCN are reported by Salom\'e
et al (2008) and CN, HCO$^+$ and C$_2$H by Bayet et al (2011).

Interpenetration of the cold filaments by the surrounding hot gas
represents another energy loss process for that gas apart from
X-radiation.  A possible conclusion from this is that gas may be
cooling from the hot phase of the intracluster medium at a higher rate
than deduced from X-ray spectroscopy alone (i.e. from radiative
cooling just in the X-ray band). This is due to some cooling of the
inner hot gas occurring non-radiatively by mixing with the cold gas,
with the energy emerging in the far infrared to UV bands. The cold gas
then hangs around for Gyrs as a slowly-accumulating reservoir of cold
molecular dust clouds, forming stars slowly and sporadically.

Molecular filaments, probably due to ram-pressure stripping, are also
seen around some galaxies in the Coma and Virgo clusters (e.g. Dasyra
et al 2012). Studies of these filaments should help our understanding
of the filaments around BCGs and vice versa.

\subsubsection{Turbulence in cool cores}

Direct measurements of the level of turbulence have been made from
X-ray line widths using the XMM-Newton RGS (Sanders et al 2010b). This
is a slitless spectrometer which looks at a region about one arcmin
across. Some luminous cool core clusters above redshifts of 0.1 appear
pointlike to this instrument, enabling the full spectral resolution to
be obtained. Several clusters including A1835 show a dozen narrow iron
and OVII lines from which velocity broadening of the X-ray coolest
components in the hot gas can be measured, yielding values less than
$300\kmps$. Turbulent energy density is then less than 10 per cent of
the thermal energy density. This is consistent with some simulations
(Vazza et al 2009). Despite AGN activity pumping out $10^{45}\ergps$ of
mechanical power, the gas flows are modest.

Indirect measurements have also been made.  A search for departures
from hydrostatic equilibrium when comparing X-ray and optically
derived gravitational potentials allows for little additional pressure
from turbulence (Churazov et al 2008). Evidence for resonant
scattering of FeXVII lines in the X-ray spectrum of several elliptical
galaxies (Werner et al 2009) limits turbulence to less than 5 per cent
of thermal values. The feedback is surprisingly gentle.

\subsection{The evolution of cool cores in clusters}

No evolution in cool core properties is seen in clusters out at $z\sim
0.5$ (Bauer et al 2005; Hlavacek-Larrondo 2012). Beyond that redshift
however, Vikhlinin et al (2007) found no cool cores in the 400 square degree
ROSAT survey. Santos et al (2010) find some in other samples but no
strong cool cores (ie with short cooling times), other than one at
$z\sim 1$ (Santos  et al 11). Can this mean rapid
evolution?
 
The lack of observed strong cool cores may be a selection effect
(Russell et al 2012) and due to enhanced AGN activity in the BCG.
Generally the central AGN in BCGs is quite sub-Eddington ($\lambda\sim
10^{-3}- 10^{-2}$). The luminous low redshift quasar H1821+643 at
$z=0.3$ in the centre of a rich cluster is a counter-example (Russell
et al 2010). The surrounding gas is in the same state as other cool
core BCGs, including a large mass of cold molecular gas (Aravena et al
2011), indicating that a powerful quasar and a cool core can
co-exist. The black holes in BCG at $z\sim 1$ may be more active than
at low redshift. Distant cool core clusters hosting central quasars
are therefore likely. If such an object is found in an X-ray survey
operating at low spatial resolution (e.g. ROSAT at 20~arcsec), and
identification is by optical spectroscopy of the brightest galaxy or
object in the error box, then the object will be optically identified
with the quasar alone if broad lines are seen in the spectrum.  The
underlying cluster could remain unnoticed if no subsequent high
spatial resolution X-ray (or optical/infrared) deep observation is
made of the quasar.

3C186 (Siemigiowska et al 2010) and PKS1229 (Russell et al 2012) are
two examples where there appear to be strong cool cores surrounding
quasars at $z\sim 1$. Lower redshift examples are presented in
Crawford \& Fabian (2003).

Until such potential selection effects are investigated further it is
difficult to speculate from observation about the evolution of cool
cores. There are of course many radio-loud quasars and radio galaxies
surrounded by extensive Ly-$\alpha$ nebulosities (McCarthy et al 1995,
Stockton, Fu \& Canalizo 2006).

Cool cores appear to be fairly robust to cluster mergers. Observations
of some merging clusters show displaced cool cores (A\,168, Hallman \&
Markevitch 2004; A\,2146, Canning et al 2012). Some simulations also
support survival (Poole et al 2006). The lack of a cool core in the
Coma cluster, for example, has long been attributed to a merger
(Fabian et al 1984). What may matter most is when the merger history
of a cluster, with early mergers being the most destructive of a cool
core (Burns et al 2008).

\subsection{The most luminous clusters}
The most X-ray luminous cluster known, RXJ\,1347-1145 with a bolometric
X-ray luminosity of $10^{46}\ergps$, is a cool core cluster, as are
many other highly luminous clusters (e.g. A1835, Zw\,3146,
RXCJ\,1504.1-0248 and several MACS clusters, Hlavacek-Larrondo et al
2012). The feedback power in these objects exceeds $10^{45}\ergps$, so
is comparable to the output of a quasar, yet their nuclei are not
exceptionally bright. Much of the energy flow is essentially
invisible. These objects are remarkably radiatively inefficient, in
the accretion flow, the jet acceleration process and the jets
themselves.  In order that the accretion flow can be so inefficient,
they probably have to operate in the ADAF regime which requires the
accretion rate to be less than about one per cent of the Eddington
rate. This in turn implies that the central black holes are
ultra-massive, well exceeding $10^{10}\Msun$ (Hlavacek-Larrondo et al
2011). 

Since the Bondi accretion rate $\dot M_{\rm Bondi}$ scales as the BH
mass squared, such high masses may make Bondi accretion feasible in
even the most luminous objects.  $\dot M_{\rm Bondi}$ also scales
inversely as the temperature of the gas at the Bondi radius to the
power 2.5 (for a given pressure), so some lower temperature gas
(e.g. at 0.5keV) associated with a weak cooling flow below 1~keV can boost
the rate further.

As well as A1835 mentioned earlier, both RXCJ\,1504.1-0248 (Ogrean et
al 2010) and MACS\,1931.8-2634 (Ehlert et al 2011) have $100-200
\Msunpyr$ of star formation, as deduced from the copious excess blue
light seen. The behaviour of feedback in these luminous clusters
appears similar to that in more typical clusters which are one or two
orders of magnitude less luminous, meaning that the processes involved
are robust. Without any feedback, radiative cooling would lead to mass
cooling rates of thousands of $\Msunpyr$ in these objects.

\subsection{Hot gas in groups and elliptical galaxies}

Dropping in X-ray luminosity by 3--4  orders of magnitude from the most
luminous clusters are elliptical-dominated groups of galaxies and
about an order of magnitude lower are individual elliptical
galaxies. Feedback can be seen operating in many of them. Most X-ray
luminous groups have cool cores with short central radiative cooling
times ($<1$~Gyr) and low central entropy (Rasmussen \& Ponman 2009;
Sun et al 2009).  A full range of bubbling behaviour is seen in these
objects (see e.g. Fig.~4).

Nulsen (2007, 2009) study a sample of 104 elliptical galaxies with
diffuse X-ray emission and find cavities in 24. Most of the objects
with cavities appear to have a heat input well in excess of their
cooling luminosity (Fig.~5). The authors suggest that the duty cycle
of bubbling is then low, at around 25 per cent, which is a sharp drop
from the situation in clusters and groups. At face value this
indicates that the duty cycle of bubbling drops with luminosity below
the high value ($\sim 100$\%) in groups and clusters.  Bubble merging,
if it occurs, could explain the higher power inferred from this
sample, but does not explain why no bubbling is detected at all in
some objects, although some of the selection effects mentioned below
may be relevant. Diehl \& Statler (2008), in a study of Chandra X-ray
data of 54 elliptical galaxies, report that the gas is almost always
assymmetrically disturbed and that this correlates with X-ray and
radio measures of AGN activity. 

More work is needed to explore all that is happening here and to
firmly decide whether the activity scales simply with host mass and
luminosity, or not. Several selection effects become important for the
detection of low power bubble activity in lower mass galaxies,
including: a) bubble size: small bubbles will be hard to resolve, even
using the sub-arcsecond resolution of Chandra, b) low X-ray surface
brightness: this can range from many tens of counts per pixel in a
long observation of a bright cluster to just a few counts per pixel in
a typical observation of an elliptical galaxy, so distinguishing a
10--20 percent drop over a small number of pixels is difficult, c) low
mass X-ray binaries produce a mess of point sources in elliptical
galaxies which need to be removed in order to distinguish the hot gas
emission and d) the gas temperature will be lower in lower mass
objects, shifting the bulk of the emission from hot gas to below the
Chandra window (effectively 0.5--7~keV).  Finally, the increasing
onset of line radiation makes the cooling function rise steeply from
$10^7$ to $10^6\K$, making it difficult to stably heat and maintain a
static atmosphere below $10^7\K$.

As an example, consider the elliptical galaxy, NGC\,720, for which the
temperature, gas mass and cooling time profiles are shown in Fig.~2
(from Humphrey et al 2010). It shows no clear central activity beyond a
weak radio source. Bubble size scales roughly as $L_{\rm cool}^{0.5}$ and
should be approximately 200~pc in radius ($\sim 2$~arcsec) for
NGC\,720. The current Chandra data on this galaxy, which is of low
surface brightness, cannot constrain the presence of any bubbles of
that size.
   
Kinetic AGN feedback may operate in any object with a hot extended
corona. Whether there is a lower cutoff in galaxy mass to such a
corona is not yet clear. Mulchaey \& Jeltema (2010) find that
elliptical field galaxies with infrared luminosity $L_K<L_*$ are
mostly devoid of hot gas. Most extended soft X-ray emission seen from
lower mass galaxies and spiral galaxies has been attributed to a
galactic fountain or outflow. Anderson \& Bregman (2011) state that
``no hot halo has been detected around a disk galaxy out to a radius
of more than a few kpc'' before reporting the detection of 40~kpc
extended emission, interpreted as a hot halo, around the massive
spiral galaxy NGC\,1961. Mass could be the important criterion here.

Recent detailed work using Hubble Space Telescope surface photometry
(Kormendy et al 2009) and integral-field spectroscopy (SAURON; Davies
2011 and references therein) is changing our picture of the internal
structure of elliptical galaxies. X-ray emitting gas is common in the
massive, slowly rotating ellipticals with inner cores. AGN feedback
acting in maintenance mode keeps much of the hot gas from cooling
(Kormendy et al 2009). The situation for the less massive, normal
ellipticals, which show excess light in the core and are rotating and
disky, is less clear, as is the origin of the structural
differences. The above selection effects may be relevant here.

Many elliptical galaxies orbit in the cores of groups and clusters of
galaxies where extensive gaseous haloes have been ram-pressure
stripped away.  These objects do have weak central X-ray nuclei with
($10^{38}<L_{\rm X}<10^{40}\ergps$ (Santra et al 2007, Gallo et al
2009). An interesting subset of these possess a minicorona, which is a
sharp-edged puddle of gas at the galaxy's virial temperature ($kT\sim
1\keV$) with a radius of 1--3~kpc (Vikhlinin et al 2001; Sun et al
2007; Santra et al 2007).  The minicorona gas probably originates as
stellar mass loss.

The Bondi accretion radius is resolved in Chandra X-ray images of some
of the nearest objects, such as M87 (Di Matteo et al 2003), enabling
the temperature and density profiles to be estimated and thus the
accretion rate determined. For other nearby elliptical galaxies the
density and temperature profiles can be extrapolated inward from
measurements made in the inner kpc. Allen et al (2006) have studied a
sample in which the kinetic power can be estimated from bubbles and
compare this with the Bondi accretion rate. A correlation emerges
indicating that a few per cent of the rest mass energy of the
accretion flow is released as mechanical energy in the jets.

\subsection{The Kinetic luminosity function}

Luminosity functions of the power radiated by quasars and AGN in
general can be readily made from careful observations of large samples
of objects. Less straightforward is to compile a kinetic luminosity
function. It has been done however by Merloni \& Heinz (2008),
Cattaneo \& Best (2009) and Mocz
et al (2012). The accretion history is inferred from the radiant
luminosity functions, and  some assumption is made connecting that to
the kinetic power history. The mass function of black holes provides
an integrated check on accreted mass. 

To relate the kinetic and radiated power, a typical scheme might be to
assume that above about one per cent of Eddington luminosity the
accretion flow is mainly radiatively efficient, with a  probability
(say 10 per cent) that jets are also present. Below that luminosity
the flow is advection dominated so the radiated power drops as the
square of the accretion rate, with the bulk being kinetic power carried
by jets. The net result of these calculations is that about half a per
cent of the accretion power emerges as kinetic energy.
 
\section{Baryon profiles at different mass scales and AGN feedback}

Early predictions for the relation between the X-ray luminosity  and
temperature of intracluster gas indicated $L\propto T^2$, based on gas
falling into dark matter potential wells of different total
masses (Kaiser 1986). This is the pure gravity prediction.  Observations show
otherwise with a relation closer to $L\propto T^{2.7}$ for clusters
with temperatures in the range of 3.5--10~keV (Markevitch 1998). It may
flatten toward the gravity prediction at higher temperatures. Below
about 2~keV there appears to be a large spread in the luminosity at
a given temperature. Some extra energy is required. 

The most likely source of the energy to heat groups is AGN (Wu, Fabian
\& Nulsen 2000, Valageas \& Silk 1999; MacCarthy et al 2011). A
significant fraction of the total power from all the black holes within
a cluster or group is required here, not just that from the central
galaxy.

The gas fraction (baryon mass vs total mass) rises outward in clusters
and approaches the cosmic value toward the virial radius
(e.g. Vikhlinin et al 2006; Allen et al 2008). Lower temperature
clusters and groups have lower gas fractions in the core which
indicates significant energy injection. (Gonzalez et al (2007) find an
increasing stellar fraction with decreasing group/cluster mass,
peaking below $10^{14}\Msun$ where stellar and gas masses are
equal. The trend is confirmed by later studies (e.g. Giodini et al
2009) but with a reduced stellar contribution (see Balogh et al 2008
for theoretical limits on the stellar contribution).) They also
approach the cosmic value at the virial radius (Humphrey et al
2012). Putting energy into intracluster or intragroup gas causes the
gas to expand, reducing its density by a much larger factor than the
temperature rises and thus {\em reduces} its X-ray luminosity. Just
how much energy has been injected depends upon when it happened. If
the injection was after the group or cluster was formed, then about
1--3~keV per nucleon is required (Wu, Fabian \& Nulsen 2000, Lapi et
al 2005), which clearly will have a major impact on low temperature
clusters and groups. Less is required if it was injected early into
gas which later fell into the cluster, since raising the adiabat
(increasing the entropy) means that the gas is more difficult to
compress. These possibilities can be discriminated against by
comparing the gas mass fraction at a fiducial radius (e.g. $R_{500}$,
Vikhlinin et al 2006; Dai et al 2010) as a function of mass and
redshift. Young et al (2011) find from this approach that continual
feedback is preferable to pre-heating.

\subsection{Powerful Radio Galaxies}

Powerful radio galaxies were relatively common at high redshifts above
$z\sim 1$ (Miley \& De Breuck 2008). These Fanaroff-Riley Class II
objects have lobes which extend many 100s kpc from the active nucleus
and, due to extensive losses (synchrotron cooling, adiabatic expansion
and inverse Compton scattering on the Cosmic Microwave Background, CMB) are
only readily observed in the radio band when they are young (Blundell
\& Rawlings 1999). The lobe energies estimated from radio  and
X-ray (Erlund et al 06) observations are high, ranging from $10^{60}-
10^{62+}\erg$ with a large uncertainty in the energy stored in protons,
which should increase the total energy. Given that the thermal energy
content of the gas in a small group of total mass $5\times
10^{13}\Msun$ is about $10^{61}\erg$, ($E_{\rm th}= 2\times
10^{61}(M_{\rm gas}/5\times 10^{12}\Msun)(kT/1\keV)\erg$) it is clear
that such giant radio galaxies can play an important role in the
evolution of intragroup gas.

It is possible that most massive galaxies have at least one outburst
of jet activity, lasting $\sim 10^8\yr$, leading to a giant radio
source in their lifetime, probably between $z\sim 1.5-3$ in the quasar
era. Estimates of the volume permeated by both ``live'' and ``dead''
radio lobes at that time can be several to tens of per cent
(Gopal-Krishna \& Wiita 2001; Mocz, Fabian \& Blundell 2011) of the
volume occupied by galaxy-forming filaments (Cen \& Ostriker 1999),
depending on the jet lifetime. Given that they occur around the most
massive galaxies which occur in proto-groups and clusters, such radio
lobes should have a significant destructive and heating effect on the
gas content.

MS\,0735.6+7421 may be a lower redshift example of a powerful radio
galaxy. Observations of distant examples are hindered in the radio by
steep Compton losses from CMB which scale as $(1+z)^4$, thus rendering
the observed radio lifetime short. 

\subsection{Similarities with  Galactic Black Hole Binaries}

A simple picture emerges in which a massive black hole in a galaxy
turns into an Eddington-limited quasar, blows away the surrounding
gas, truncating both further star formation in that galaxy and quasar
activity. It may even involve a giant radio outburst.
The galaxy dies and any later infalling or cooling gas is
heated by jetted, maintenance-mode feedback (e.g. Churazov et al
2005). The behaviour of the black hole resembles that of outbursts in
Galactic Black Hole Binaries (BHB). These objects consist of a stellar
mass black hole ($4-20\Msun$) in close orbit about a normal
star. Accretion instabilities cause the accretion rate to increase to
the Eddington rate then drop by several orders of magnitude. Although
they are not important sources of feedback in a galaxy, some BHB do
blow bubbles (e.g. Gallo et al 2005). 

Outbursts follow a common pattern (Fender \& Belloni 2004; Remillard \&
McClintock 2006); the source beginning in the weak radio-emitting low
state, rising in luminosity in all bands towards the Eddington limit
where its spectrum softens and the radio emission peaks with a strong 
outburst. Radio emission then ceases until the luminosity drops
 and the spectrum hardens into the low state again. As the
source goes into a quiescent state the flow becomes advection
dominated.

Much of black hole accretion including timescales and luminosities
scale with mass. The radio emission scales in a more complex manner
along a fundamental plane (Merloni, Heinz \& Di Matteo 2003), but the
general pattern may remain for AGN. The timescales are far too long
for us to watch an outburst and their behaviour must be pieced
together from the populations we observe. There are some plausible
connections however, with the luminous outbursts correlating with
powerful FRII sources and the weaker low state with FRI sources.

A further property of BHB which may prove important is that winds and
jets appear to anticorrelate (Miller et al 2008; Neilsen \& Lee 2009;
King et al 2012; Ponti et al 2012). Winds occur at high Eddington
fractions when jets are not observed, and vice versa at low Eddington values.
  
Atomic processes do not scale for accretion flows, so
ionization-dependent accretion instabilities may not be relevant in
quasars, and outbursts may be triggered instead by mergers. Clear
evidence for this has yet to be established. Submillimetre galaxies at
high redshifts often show evidence for strong disturbances which could
be mergers, as do some nearby low-luminosity AGN (Sec.~2.5), but most
AGN at intermediate redshifts show no more evidence for merging than
do control non-active galaxies.

There are many uncertainties in the evolution of AGN, apart from the
role of mergers. Does a massive galaxy undergo a single quasar phase
or several? Is the observed luminous phase preceded by a Compton thick
phase? How representative are observations of low redshift AGN to
conditions around powerful quasars at $z\sim 2$?  How much do jets
depend on the spin of the black hole and how does the spin evolve? Is
a cluster cool core shaped by a powerful outburst early on and 
maintained thereafter by black hole spin, which is kept ``topped up''
by sporadic accretion?

\section{Future studies}

Contributions to understanding AGN feedback can be expected from all
wavebands, but there are some clear advances which can be anticipated
from instruments and telescopes being built or planned for the next
few years.  In particular, the JAXA/NASA/ESA X-ray observatory ASTRO-H
(Takahashi et al 2010), to be launched in 2014, will offer
non-dispersive high spectral resolution X-ray spectroscopy on a
spatial scale of 1.5 arcmin using a microcalorimeter. It will reveal
the number and nature of outflows in quasars and other AGN through
absorption spectroscopy, being sensitive to all ionizations stages of
iron, for example. It will also, for the first time, map velocity
flows of hot gas in galactic outflows, and the hot atmospheres in
elliptical galaxies, groups and clusters. Since much of the energy
feedback in those objects is mechanical and though the motion of gas,
this will take our understanding to the next level of detail.

Sensitive, low-frequency radio observations at good spatial
resolution, as expected from LOFAR and other planned radio
telescopes leading to the SKA (Square Kilometre Array) will reveal the
old electron populations in the bubbles of cluster cores and giant
radio galaxies. Maps of the low-frequency radio Sky at redshift 2
should reveal just how much and where feedback from powerful FRII
sources has occurred.

At GHz frequencies the Jansky Very Large Array (JVLA) will map the
interactions of jets with surrounding plasma, and reveal the magnetic
field structure of the plasma through much more detailed Faraday
Rotation observations.  

\begin{figure}
\begin{center}
 \includegraphics[width=3.5in]{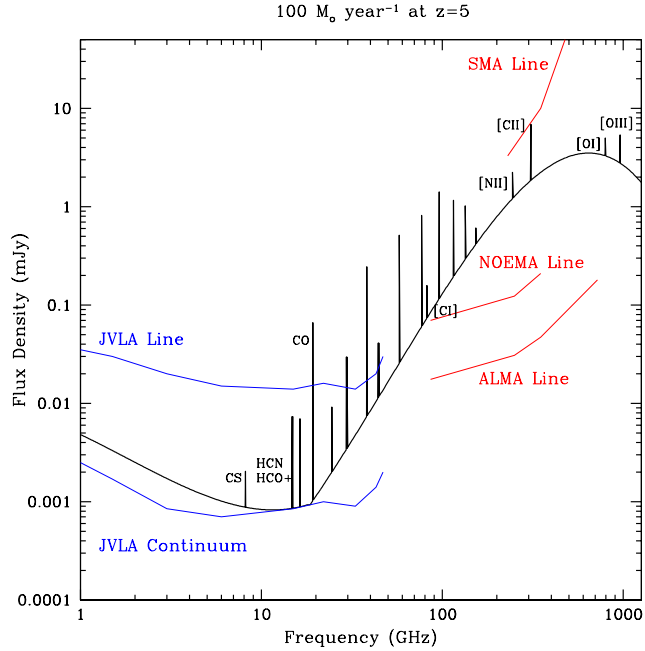}
 \vspace*{0 cm}
 \caption{Radio through submm spectrum of a galaxy forming stars at a
   rate of $100\Msunpyr$ redshifted to $z=5$. The line and continuum
   sensitivity for the JVLA in 12 hr and the line sensitivity for ALMA
   and existing submm interferometers are shown (adapted from Carilli
   et al 2011). NOEMA is an ongoing expansion of the IRAM Plateau de
   Bure interferometer by a factor of two in collecting area. SMA is
   the Submillimeter Array in Hawaii. }
   \label{fig1}
\end{center}
\end{figure}

The role of radiative feedback needs high spatial and spectral
resolution observations at wavelengths that can penetrate the
inevitable high absorption present. ALMA (the Atacama Large
Millimetre/submillimetre Array) will be a leader in detecting and
resolving molecular and dust components and their motions. For high
redshift objects where feedback may be a dominant process, a
combination of the the JVLA, which will measure the radio continuum to
distinguish AGN and star formation, low order CO transitions to give
total molecular gas mass, dynamics and low order dense gas tracers
(HCN, HCO$^+$) to probe dense star cores, and ALMA, which will obtain
the excitation from high order CO, the dust continuum and star
formation and fine structure lines to give the cooling rates, further
distinguish AGN and star formation and measure dynamics, will be
crucial (Fig.~10: Carilli et al 2011).

The James Webb Space Telescope will observe the rich, rest-frame,
optical band in distant objects, enabling many powerful diagnostics to
be used. AGN feedback will be an important goal for the next generation
ground-based optical and near infrared telescopes.

Both large-area surveys (e.g. Sloan Digital Sky Survey, FIRST- Faint
Images of the Radio Sky at Twenty centimetres) and single-object
studies (e.g. Mrk\,231, APM\,08279+5255, the Perseus cluster, M87 etc.)
will continue to be essential.  Very significant advances in our
observational understanding of AGN feedback can be confidently
expected in the present decade.

\section{Summary}

An active nucleus interacts with the gas in its host galaxy through
radiation pressure, winds and jets. The consequences can be profound
for the final mass of the stellar component of the galaxy as well as
for the black hole. There is clearly enough energy and momentum
produced by the AGN to expel the interstellar medium of the host
galaxy. How, when and if it does so are the important questions.

It appears that the radiative or wind mode was most active when the
AGN was a young quasar. At that stage the galaxy had a large component
of cold molecular gas and the nucleus was probably highly
obscured. Obscuration has meant that direct observational evidence is often
circumstantial, relying on nearby analogues.  Progress has been
difficult and slow, but is expected to accelerate soon following
observations with ALMA, JVLA and other telescopes.

The kinetic mode on the other hand is more easily observed, albeit at
X-ray and radio wavelengths, since it is acting now in nearby massive
objects. The surrounding gas is hot, highly ionized and mostly
transparent. Although the gross energetics are roughly understood, the
details are not. Bubbling of jetted energy from the central black hole
appears to scale well over 3 to 4 orders of magnitude in luminosity
from the most luminous clusters to small groups. The behaviour in
individual elliptical galaxies and bulges is uncertain, partly because
it is more difficult to detect. Non-radiative cooling of hot gas by
mixing with cold gas may be an important link in the process.    

An attractive possibility is that the radiative mode shaped the
overall galaxy and black hole mass at early times and the kinetic mode
has since maintained that situation where needed (Churazov et al
2005). 

Powerful giant radio outbursts back at $z\sim 1.5-3$ in all massive galaxies
may have been common, heating and shaping the gas not only in the host
galaxy but in the host groups and protoclusters.

Observational evidence is growing that the baryonic part of the low
redshift Universe has been shaped by the energy and momentum output of
black holes, through AGN feedback. This has profound implications for
our understanding of galaxy, group and cluster evolution and has
ramifications for precision cosmology using galaxies (van Daalen et al
2011). AGN feedback appears to be an important aspect in the
complexity of the baryonic universe.

\section{Acknowledgements}
I thank Becky Canning, Chris Carilli, Julie Hlavacek-Larrondo, Fill
Humphrey, Brian McNamara, Jeremy Sanders, Greg Taylor and particularly
John Kormendy, for help and advice.  

\begin{table}
  \caption{Observational Evidence for AGN Feedback}
\centering
\begin{tabular}{ll}
  	\hline
High velocity broad absorption lines in quasars & strong\\
Strong winds in AGN  & strong\\
1000 km/s galactic outflows & strong \\
Bubbles and ripples in BCGs & strong\\
Giant radio galaxies & strong\\
Lack of high SFR in cool cluster cores & indirect \\
$M-\sigma$ relation & indirect\\
Red and dead galaxies & indirect\\
Lack of high lambda, moderate $N_H$, quasars & indirect\\
Steep $L-T$ relation in low $T$ clusters and groups & indirect\\
        \hline
\end{tabular}

\label{par.tab}

\end{table}


\end{document}